\documentclass[apj]{emulateapj}
\usepackage{apjfonts}
\usepackage{graphicx}
\usepackage{epsfig}
\usepackage{amsmath}
\usepackage{amssymb}
\usepackage{amsfonts}
\usepackage{natbib}          %  This is required to use the apj.bst
\pdfoutput=1

\shorttitle{Anomalous Microwave Emission in HII regions}
\shortauthors{Paladini et al.}

\slugcomment{accepted to ApJ on September 11th 2015}

\begin{document}

\title{Anomalous Microwave Emission in HII regions: is it really anomalous ? The case of RCW 49}

\author{
%Paladini\altaffilmark{1}, 
%et al.
Roberta Paladini\altaffilmark{1},
Adriano Ingallinera\altaffilmark{2},
Claudia Agliozzo\altaffilmark{2,3},
Christopher T. Tibbs\altaffilmark{4},
Alberto Noriega-Crespo\altaffilmark{5},
Grazia Umana\altaffilmark{2},
Clive Dickinson,\altaffilmark{6}
Corrado Trigilio\altaffilmark{2}
}

\altaffiltext{1}{
Infrared Processing Analysis Center, 
California Institute of Technology, 
770 South Wilson Ave., 
Pasadena, CA 91125, USA
}
\altaffiltext{2}{
Osservatorio Astrofisico di Catania, 
Via S. Sofia 78, 
95123 Catania Italy
}
\altaffiltext{3}{
Departamento de Ciencias Fisicas, Universidad Andres Bello, 
Avda. Republica 252, Santiago  8320000, Chile
}
\altaffiltext{4}{
Scientific Support Office,
Directorate of Science and Robotic Exploration,
European Space Research and Technology Centre (ESA/ESTEC),
Keplerlaan 1, 2201 AZ,
Noordwijk, The Netherlands
}
\altaffiltext{5}{
Space Telescope Science Institute,
3700 San Martin Drive,
Baltimore, MD 21218, USA
}
\altaffiltext{6}{
Jodrell Bank Centre for Astrophysics, 
Alan Turing Building, 
School of Physics \& Astronomy, 
The University of Manchester, 
Oxford Road, Manchester M13 9PL, UK
}
	
\begin{abstract}
The detection of an excess of emission at microwave frequencies with respect to the predicted free-free emission has been reported 
for several Galactic HII regions. Here, we investigate the case of RCW 49, for 
which the Cosmic Background Imager tentatively ($\sim$ 3$\sigma$) detected Anomalous Microwave
Emission at 31 GHz on angular scales of 7'. Using the Australia Telescope Compact Array, we carried out a multi-frequency (5 GHz, 19 GHz and 34 GHz) 
continuum study of the region, complemented by observations of the H109$\alpha$ radio recombination line. The analysis shows that: 
1) the spatial correlation between the 
microwave and IR emission persists on angular scales from 3.4' to 0.4'', although the degree of the correlation slightly decreases at higher frequencies and on  
smaller angular scales; 2) the spectral indices between 1.4 and 5 GHz are globally in agreement with optically thin free-free emission, however, $\sim$ 30$\%$ 
of these are positive and much greater than -0.1, consistently with a stellar wind scenario; 3) no major evidence for inverted free-free radiation is found, indicating  
that this is likely not the cause of the Anomalous Emission in RCW 49. Although our results cannot rule out the spinning dust hypothesis to explain the tentative detection of 
Anomalous Microwave emission in RCW 49, they emphasize the complexity of astronomical sources very well known and 
studied such as HII regions, and suggest that, at least in these objects, the reported excess of emission 
might be ascribed to alternative mechanisms such as stellar winds and shocks.
\end{abstract}


\section{Introduction}

Recent years have brought direct evidence that an additional
component of emission is present both in the diffuse Interstellar Medium (e.g. Davies et al. 2006) and in individual sources on or close to the Galactic Plane 
(e.g. Planck Collaboration XX, 2011; Planck Collaboration Int. XV, 2014). This component manifests itself in the microwave regime
(10 - 60 GHz) and is often denoted with the term {\em{anomalous}}, as it
appears to be closely correlated with Infrared (IR) data, therefore suggesting a possible association
with dust, despite exhibiting a pronounced excess with
respect to the predicted thermal vibrational dust emission at these frequencies. Draine $\&$ Lazarian
(1998a) have proposed that very small
rotating dust grains (Polycyclic Aromatic Hydrocarbons, i.e. PAHs, or Very Small Grains, i.e. VSGs) are 
responsible for the observed microwave excess. 

The Cosmic Background Imager (hereafter CBI, Padin et al. 2002), observing the sky at 31 GHz and with a synthetic beam of
$\sim$ 6.8',  detected 
Anomalous Microwave Emission (AME) at a 3.3$\sigma$ level in the HII region RCW 49 (Dickinson et al. 2007). The prediction of free-free emission for the 
source was based on a power-law fit of continuum data in the range 2.7 -- 15 GHz, providing an expected 31 GHz flux of 99.6 $\pm$ 13.4 Jy, i.e. much lower 
than the flux measured by CBI of 146.5 $\pm$ 5.2 Jy. Signatures of AME were also reported by CBI for RCW 175 which, with a 14.3$\sigma$ detection (Dickinson et al. 2009; Tibbs et al. 2012), represents 
to date the best example of microwave excess emission in an HII region, as well as 
for 6 other Southern HII regions, and by the Very Small Array (VSA)   
in nine HII regions in the Northern hemisphere (Todorovic et al. 2010).{\footnote{For a comprehensive summary of AME observations 
in HII regions, we refer the reader to Dickinson (2013).}}  Despite these claims for a detection, the observational scenario appears far 
from clear: a sample of sixteeen HII regions observed by Scaife et al. 2008 
showed no statistically significant excess, and the original claim of an AME 
detection in the HII region LPH+201.6+1.6 (Finkbeiner et al. 2002) was later revised by CBI (Dickinson et al. 2006) and VSA measurements (Scaife et al. 2007).  

\begin{figure*}
\label{fig:irac_cbi}
%\hspace*{-3truecm}
\begin{center}
\hspace*{-2.5truecm}
  \includegraphics[width=0.75\linewidth]{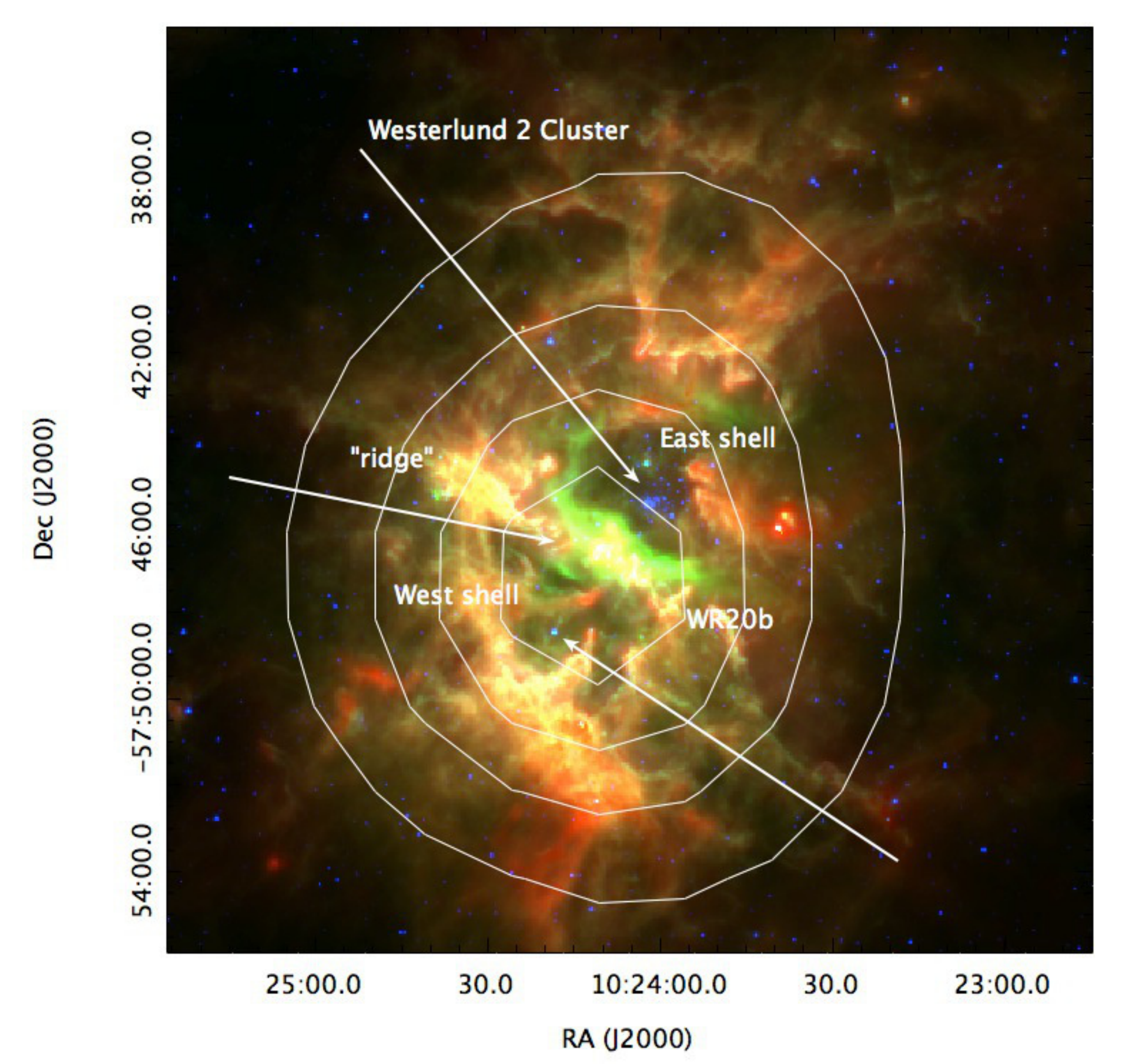}
  \caption{31-GHz CBI contours overlaid on a composite 3-color image of RCW 49: IRAC 3.6 $\mu$m (blue) and 8 $\mu$m (green) from the GLIMPSE survey, PACS 160 $\mu$m (red) from 
the Herschel OT2 program {\em{''Unveiling the misterious case of RCW 49: a powerful HII region with associated Anomalous Microwave Emission''}} (PI.: R. Paladini). The IRAC 
3.6 $\mu$m band shows the stellar content of the region, including the bright WR20b and Westerlund 2 cluster. The 8 $\mu$m band is dominated by the 7.7 $\mu$m PAH feature, 
and it highlights the Photo Dissociation Region (PDR) associated with RCW 49. Finally, the PACS 160 $\mu$m data evidence the presence of cold dust, both in the PDR and in its surroundings.}
\end{center}
\end{figure*}

In general, the possibility of detecting AME in HII regions is very intriguing from a theoretical point of view. 
On one side, HII regions are ideal candidates to harbor this emission, as processes common to this type of environment, such as 
plasma drag and photoelectric effect, can generate enough angular momentum to spin the grains and 
induce detectable microwave radiation. However, significant dust grain depletion is expected to occur in HII regions as a consequence
of the intense radiation pressure and stellar winds (Inoue 2002; Draine 2011). Even more importantly, embedded optically thick Ultra Compact HII (UCHII) regions 
could mimic, in this case, an excess at microwave frequencies. In fact UCHII regions are characterized by high emission meaures 
(i.e. 10$^{9-10}$ pc cm$^{-6}$) and by self-absorbed free-free emission at frequencies up to several GHz. 
This self-absorbed free-free could potentially generate the observed rising spectrum attributed to AME. 

Motivated by this complex scenario, in 2008 we requested Australia Telescope Compact Array (ATCA) high-resolution observations of RCW 49. The goal of 
the observations was to follow-up on the original CBI detection, in order to shed light on the nature of the reported microwave excess.   

RCW 49, located at (RA, Dec) = (10$^{h}$ 24$^{m}$ 14.6$^{s}$, -57$^{\circ}$ 46$^{\prime}$ 58''), is the most luminous and massive HII region of the Southern Galaxy (e.g. Paladini et al. 2003), with 
a bolometric luminosity of 1.4 $\times$ 10$^{7}$ L$_{\Sun}$ and a stellar mass of $\sim$ 3 $\times$ 10$^{4}$ M$_{\Sun}$ 
(Vacca et al. 1996). The source has an estimated age of 2-3 Myr (Piatti et al. 1998), while its distance remains uncertain, with different authors 
suggesting a distance anywhere between 2.3 (e.g., Brand $\&$ Blitz 1993) and 6 -- 7 kpc (e.g., Benaglia et al. 2013). RCW 49 extends for $\sim$ 2 degrees and is characterized by a very complex
morphology. At its center is the Westerlund 2 (Wd2, Westerlund 1960)
compact cluster, comprising a dozen of OB stars. An additional set of about 30 new OB star candidates is found 
in and around Wd2. Beyond the cluster core there are three more massive stars, i.e. a star of spectral type
O4 or O5 (Rauw et al. 2007), a binary Wolf-Rayet star, WR20a, which is the most massive binary system in the Galaxy
with a well-determined mass (Rauw et al. 2005), and another Wolf-Rayet star, WR20b (van der Hucht 2001), situated several arcminutes away. 

RCW 49 has been extensively investigated at both radio and IR wavelengths. In the radio continuum, data are available at 408 MHz (73 cm, Shaver $\&$ Goss 1970b), 
2.7 GHz (11 cm, Day et al. 1972), 5 GHz (6 cm, Caswell $\&$ Haynes 1987), 8.7 GHz (3.4 cm, McGee et al. 1975) and 14.7 GHz (2 cm, McGee $\&$ Newton 1981). All these 
observations are characterized by angular resolutions ranging from 8.2' (at 2.7 GHz) to $\sim$ 2' (e.g 14.7 GHz).  The Molonglo Observatory Synthesis Telescope (MOST) 
carried out sub-arcmin resolution observations at 0.843 GHz (35.6 cm, Whiteoak et al. 1989), and more recently the ATCA  
targeted the core of RCW 49 at 1.4 and 2.4 GHz (21 cm and 12.5 cm, Whiteoak $\&$ Uchida 1997,  hereafter WU97), as well as at 5.5 and 9.0 GHz (5.4 cm and 3.3 cm, Benaglia et al. 2013). 
These observations span angular resolutions from 10'' to 2''. All these data show that the core of RCW 49 is dominated by a two-shell  
system, the larger (7.3' in diameter) and more massive of which is located to the East and is centered on the Wd2 cluster, while the second shell (4.1' 
in diameter), located to the West, surrounds WR20B. 

\begin{deluxetable*}{cccccc}
\tablecaption{Observations summary}
\tablehead{
\colhead{Observation~Date} &
\colhead{ATCA~Configuration} &
\colhead{Central~Wavelength} &
\colhead{Central~Frequency} &
\colhead{Bandwidth} &
\colhead{Integration~Time} \\
\colhead{} &
\colhead{} &
\colhead{[cm]} &
\colhead{[GHz]} &
\colhead{[MHz]} &
\colhead{[hr]}
}
\startdata
2008~Dec~14-15 & 750 B     &    5.99   &  5.0089 &  16           &    10.9      \\
2009~Jan~23-24 & 1.5 C     &    5.99   & 5.0089 & 16         &     10.4       \\
2009~Jan~25 & EW 352      &    1.35  & 22.2350$^{\dag}$    &    8         &     10.1      \\
2009~Feb~7-8 & EW 352     &     0.87     &   34.5600 & 128       &      10.1      \\
2009~Feb~8-9 & EW 352      &  1.62+1.54     &  18.4960+19.5200 &  128           &       10.4     \\
\enddata
\tablecomments{$^{\dag}$ These data were not used for the analysis. See the text for more details.}
\end{deluxetable*}

At shorter wavelengths, RCW 49 has been observed with the Spitzer Telescope by IRAC, as part of the GLIMPSE survey (Benjamin et al. 2003), 
and by MIPS (PI. J. Houck, pid 63). These combined observations provide information in five photometric 
bands, from 3.6 $\mu$m to 24 $\mu$m (i.e. 3.6 $\mu$m, 4.5 $\mu$m, 5.6 $\mu$m, 8 $\mu$m and 24 $\mu$m), with a spatial resolution from 2'' to 6''. 
We note that only partial data exist at 24$\mu$m, due to hard saturation in the bright core of the source. A comprehensive analysis of the GLIMPSE data for the region is
provided in Whitney et al. (2004). This area of the sky was also covered by the Wide-field Infrared Survey Explorer (WISE, Wright et al. 2010) all-sky survey at 
3.4, 4.6, 12 and 22 $\mu$m, with spatial resolution from 6'' to 12''. As in the case of GLIMPSE, severe saturation of the core of RCW 49 occurs at 22 $\mu$m. Finally, 
RCW 49 was targeted by dedicated Herschel (Pilbratt et al. 2010) PACS (70 $\mu$m and 160 $\mu$m, at 6'' and 12'', respectively) and SPIRE (250 $\mu$m, 350 $\mu$m, 500 $\mu$m, at 18'', 25'' and 35'', respectively) 
Cycle 2 Open Time parallel mode observations (PI. R. Paladini). The analysis of these Herschel data will be the subject of a forthcoming publication (Paladini et al., in preparation).  

Noticeably, the brightest emission, at both radio and IR wavelengths, is found in correspondence of the {\em{ridge}} located between the two shells (see Figure~1). 

The paper is organized as follows. In Section~2 we describe the new ATCA continuum and radio recombination line observations of the core of RCW 49. 
In Section~3 we provide details on the data reduction. In Section~4 we analyze this new data set. 
Finally, we discuss our results in Section~5, and present the conclusions in Section~6.

\section{{\it ATCA} Observations}
\label{sec:obs}

With the ATCA multi-frequency high-resolution observations, we intended to achieve three main objectives: 

1) search for a  spatial correlation of the radio continuum emission with dust emission, and study its dependendance 
on frequency; 2) generate a spectral index map
to investigate the nature of the physical mechanism responsible for the observed microwave excess; 
3) verify the possibility that the AME deteted by the CBI experiment is due to optically thick free-free 
emission associated to UCHII regions. For 1) and 2) we needed continuum data, while point 3) can be addressed by means of 
radio recombination line (RRL) observations. 

To this end, we requested observing time at four different frequencies (5 GHz, 18/19 GHz, 22 GHz and 34 GHz) and in 
different array configurations (750-m/1.5-km, 1.5-km/EW 352 and EW 352). At 5 GHz, we observed simultaneously the continuum and the H109$\alpha$ line, whose rest-frame 
frequency is 5.0089 GHz. 
Table~1 provides a summary of the observations. Due to bad weather conditions, the 22 GHz data are of poor quality. Therefore, these data 
were not included in our analysis and are omitted in Table~2 and Table~3. 

The CBI observations covered an area which extends 7.8$^{\prime}$ $\times$ 5.6$^{\prime}$ and is centered on (RA, Dec) = (10$^{h}$ 24$^{m}$ 20$^{s}$, -57$^{\circ}$ 44$^{\prime}$ 57''). 
This area corresponds to the West shell, comprising the WR20b cluster and the {\em{ridge}} (see~Figure 1). The ATCA observations were designed to overlap with the CBI observations. 
The 18/19 GHz and 34 GHz observations were carried out in mosaic mode (with 49 pointings at 19 GHz and 63 pointings at 34 GHz), and using a 128 MHz bandwidth, 
to allow an adequate sampling of the continuum emission. The area mapped at 18/19 GHz covers the same field of view of the 5 GHz data while, due to 
partially adverse weather conditions during the observational campaign, the area mapped at 34 GHz covers only the region of the {\em{ridge}} and its 
immediate surroundings. To observe the H109$\alpha$ line, we used a 16 MHz bandwith
with 512 channels, which provides a spectral channel resolution of 1.875 km/s.

\section{Data Reduction} 

In this section we describe the generation of the 5, 19 and 34 GHz maps displayed in 
Figure~2, 3 and 4, and the extraction of the H109$\alpha$ line shown in Figure~5.

\subsection{Radio continuum data}

The MIRIAD software (Sault et al., 1995){\footnote{http://www.atnf.csiro.au/computing/software/miriad/}} was used to perform the standard calibration steps, which include: 1)
flagging corrupted data, spikes and baselines; 2) performing bandpass calibration; 3) removing bad channels; 4) improving bandpass solutions; 5)
determining time-based antenna gains by interpolation between scans taken
on the phase calibrator; 6) flux-density bootstrapping. A complete list of bandpass, flux and phase calibration sources is provided in Table~2.
In total, no more than 3$\%$ - 5$\%$ of the data were flagged.

\begin{deluxetable}{cccc}
\tablecaption{List of Calibrators}
\tablehead{
\colhead{Calibrator \hspace*{0.05truecm} Type} &
\colhead{5 \hspace*{0.02truecm} GHz} &
\colhead{19 \hspace*{0.02truecm} GHz} &
\colhead{34 \hspace*{0.02truecm} GHz}
}
\startdata
Bandpass  &  1934-638 &     0537-441           &    0537-441      \\
Flux     &  1934-638  &     1934-638         &      1934-638      \\
Phase     & 1036-52   &       1045-62      &      1045-62      \\
\enddata
\end{deluxetable}

At 5 GHz, in order to achieve the best $uv$-coverage, as well as the best S/N, we concatenated the data sets obtained in the 750 B and 1.5 C configurations.
We also concatenated the 18 and 19 GHz data sets, as we do not expect any significant changes in the spectral index between these two frequencies. Hereafter
we will only discuss results on the combined, final images.

\begin{deluxetable*}{cccccc}[h]
\tablecaption{ATCA Datasets}
\tablehead{
\colhead{Frequency} &
\colhead{Synthetic \hspace*{0.05truecm} Beam} &
\colhead{I$_{\max}$} &
\colhead{rms} &
\colhead{$\theta_{min}$} &
\colhead{$\theta_{max}$}\\
\colhead{[GHz]} &
\colhead{[arcsec]} &
\colhead{[mJy/beam]} &
\colhead{[mJy/beam]} &
\colhead{[arcsec]} &
\colhead{[arcmin]}
}
\startdata
  5 & $7.3 \times 7.3$ & 288.7 & 4.9 & 2.7 & 3.4\\
  19 & $8.5 \times 6.1$ & 175.4 & 1.2 & 0.7 & 1.7\\
  34 & $8.3 \times 4.2$ & 28.4 & 1.1 & 0.4 & 1\\
\enddata
\tablecomments{The parameters listed above for the 5 GHz map refer to the original data set, i.e. prior to filtering 
in the $uv$-plane (see Section~4.2). $\theta_{min}$ and $\theta_{max}$ indicate, respectively, the smallest and largest imaged angular 
scales.}
\end{deluxetable*}

\begin{figure*}
\label{fig:6cm_map}
 \begin{center}
 \includegraphics[width=0.6\linewidth]{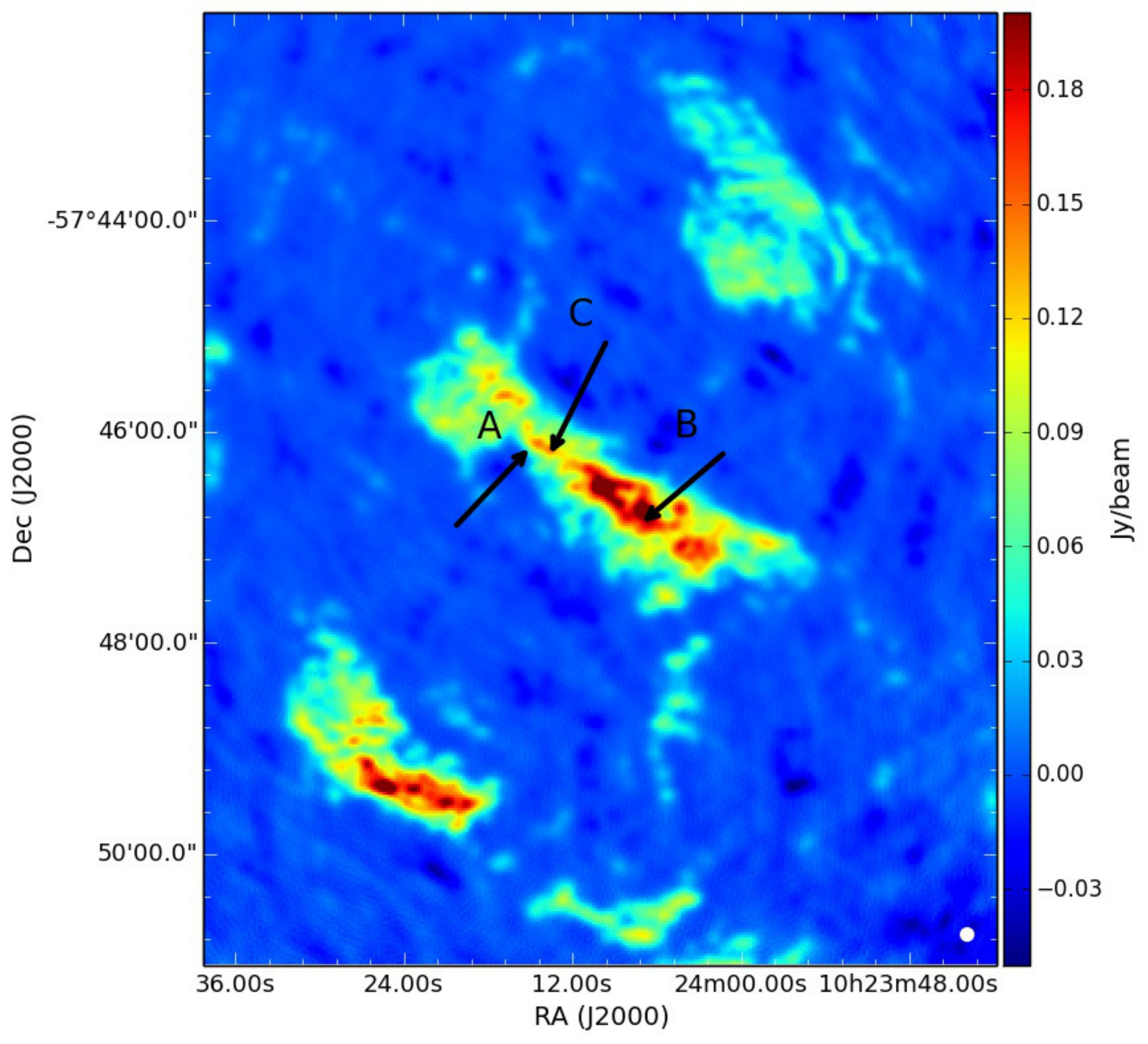}
 \caption{ATCA 5 GHz map. The synthetic beam is shown in the bottom right corner. The map units are Jy/beam.}
\end{center}
\end{figure*}

\begin{figure*}[h]
\label{fig:1e6cm_8e8mm_maps}
 \begin{center}
 \includegraphics[width=0.6\linewidth]{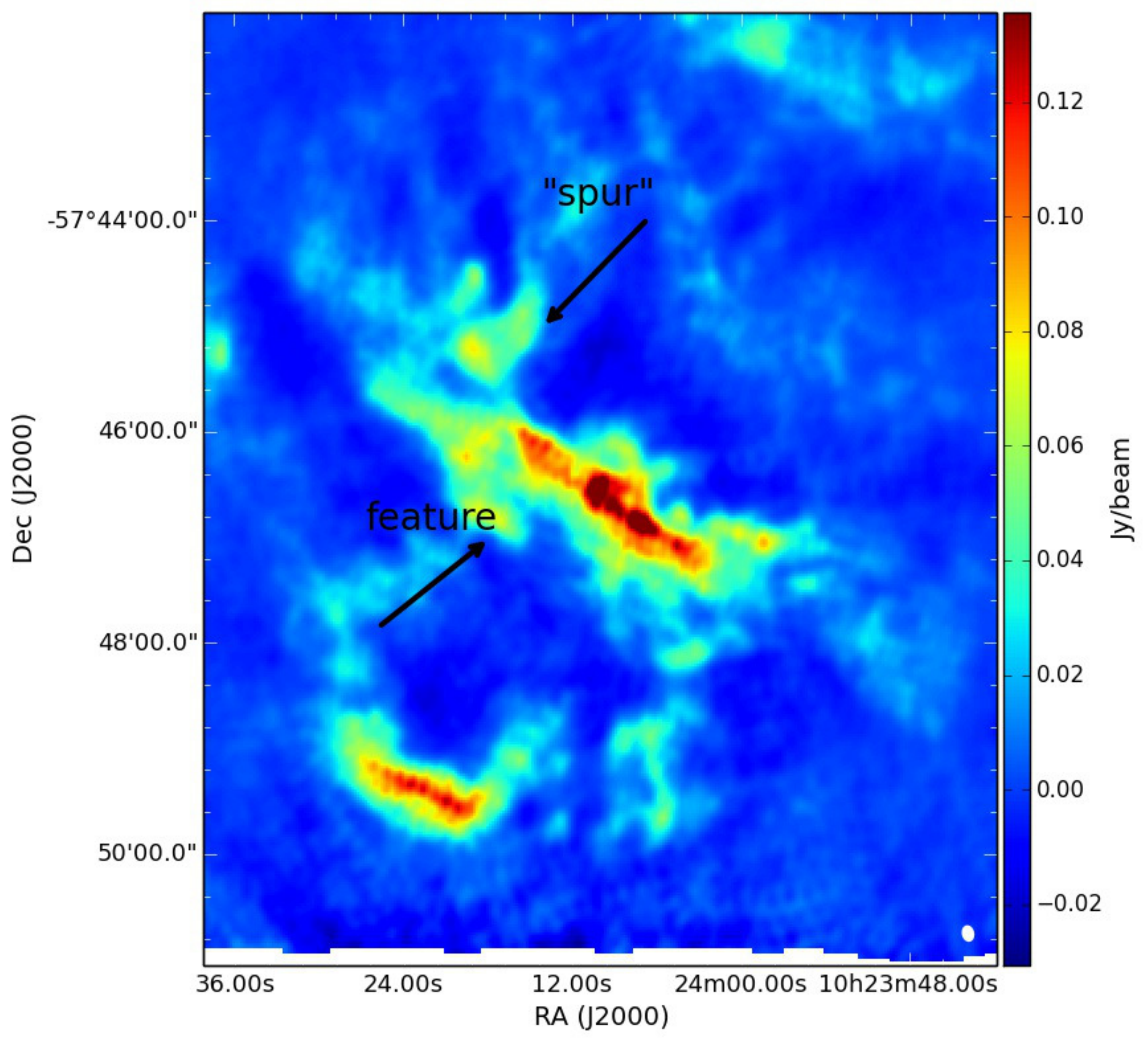}
 \caption{ATCA 19 GHz image of RCW49. The synthetic beam is shown in the bottom right corner. The beam PA is -12.3$^{\circ}$.
          The map units are Jy/beam.}
\end{center}
\end{figure*}

\begin{figure*}[h]
\label{fig:1e6cm_8e8mm_maps}
 \begin{center}
 \includegraphics[width=0.6\linewidth]{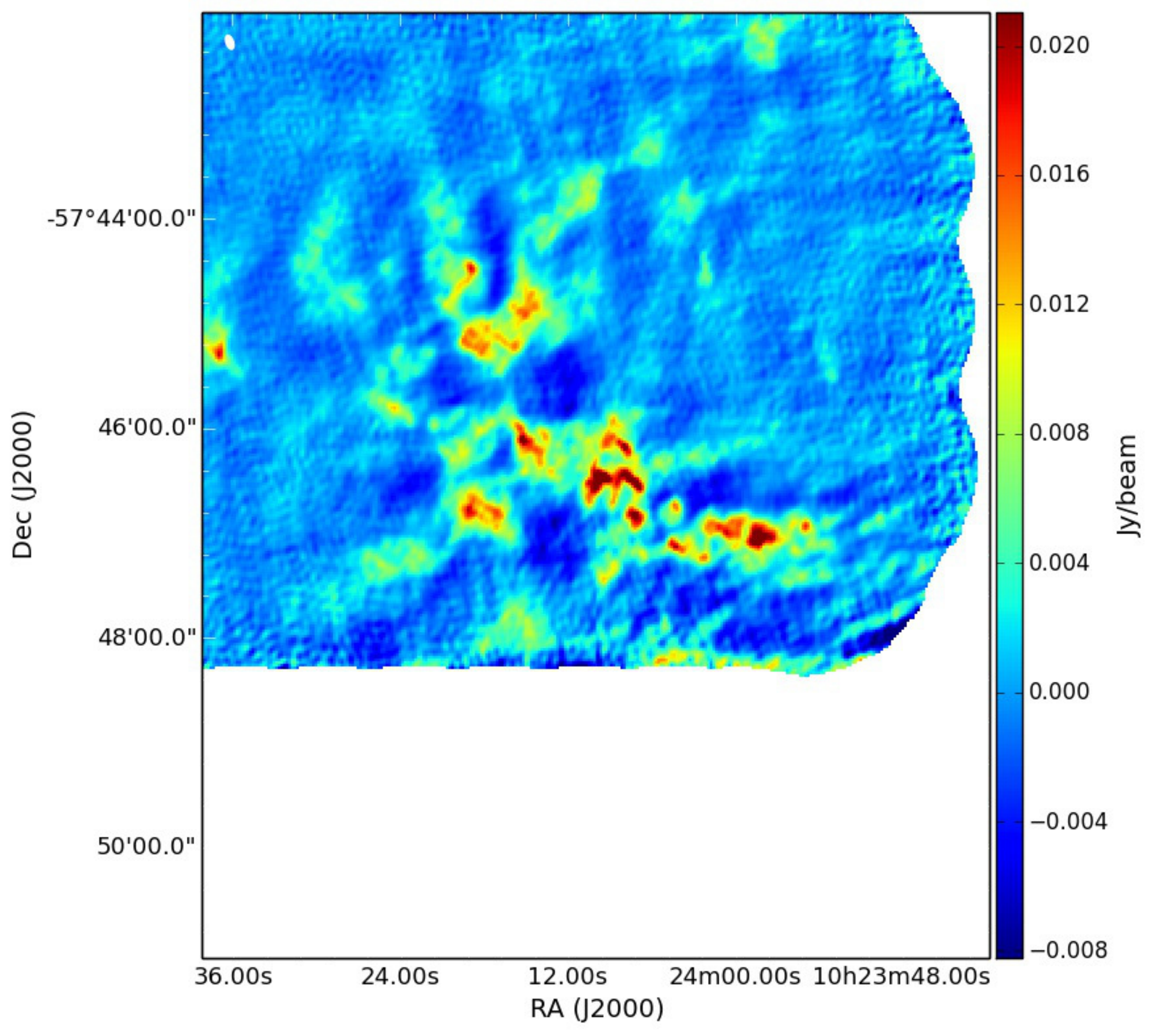}
 \caption{ATCA 34 GHz image of RCW49. The synthetic beam is shown in the top left corner. The beam PA is 160.2$^{\circ}$.
          The map units are Jy/beam.}
\end{center}
\end{figure*}

The imaging process was carried out with CASA{\footnote{http://casa.nrao.edu}} v3.3. For all frequencies, with the exception of 5 GHz, we used a Briggs weighting,
which allows a good match between noise and resolution. The parameter \texttt{robust} was set to zero, to minimize sidelobe contributions from bright sources.
At 5 GHz, we used a natural weighting scheme, in order to assign more weight to the shortest spatial frequencies,
and achieve a better sensitivity at the largest angular scales. Deconvolution of the dirty images was performed by default with
the Clark algorithm. Alternatively, the H\"ogbom algorithm was used when the $uv$-coverage was too poor for imaging. In addition, a correction for the primary beam was applied. 
Table~3 provides the synthetic beams, peak flux densities and noise of the final maps, shown in Figure~2, 3 and 4. 

\begin{center}
\begin{deluxetable*}{cccccccccccc}
\tablecaption{Recombination line parameters}
\tablehead{
\colhead{Region} &
\colhead{RA} &
\colhead{Dec} &
\colhead{T$_{L}$} &
\colhead{T$_{C}$} &
\colhead{rms} &
\colhead{SNR} &
\colhead{$\Delta V$} &
\colhead{$V$} &
\colhead{$T_{e}$} &
\colhead{EM} &
\colhead{n$_{e}$}\\
\colhead{} &
\colhead{h\hspace*{0.05truecm} m \hspace*{0.05truecm} s} &
\colhead{d\hspace*{0.05truecm} m \hspace*{0.05truecm} s} &
\colhead{[mJy/beam]} &
\colhead{[mJy/beam]} &
\colhead{[mJy/beam]} &
\colhead{} &
\colhead{[km/s]} &
\colhead{[km/s]} &
\colhead{[K]} &
\colhead{[pc \hspace*{0.02truecm} cm$^{-6}$]} &
\colhead{[cm$^{-3}$]}
}
\startdata
A        &  10 24 15.1 & -57 46 10    &  6.94  & 33.24      & 2.3 & 3.04 & $\phantom{1}9.2\pm2.6$ & $\phantom{-}16.0\pm1.8\phantom{-}$ & $\phantom{0}5750\pm\phantom{0}1960$ &   $\phantom{0}170\pm98$   & $\phantom{0}13.0\pm3.8$\\ 
B        &  10 24 07.0 & -57 46 52    &  4.26  & 21.80      & 1.6 & 2.71 & $\phantom{1}3.1\pm1.0$ & $\phantom{1}{-2.6}\pm5.5\phantom{-}$ & $\phantom{0}15700\pm\phantom{0}8430$  &  $\phantom{0}159\pm144$  & $\phantom{0}12.6\pm5.7$\\
C        &  10 24 13.6 & -57 46 13    &  8.79  & 36.90      & 3.9 & 2.28 & $\phantom{1}12.4\pm2.5$ & $\phantom{-}16.0\pm1.8\phantom{-}$ & $\phantom{0}3955\pm\phantom{0}1110$ &    $\phantom{0}166\pm80$   & $\phantom{0}12.9\pm3.1$\\
\enddata
\end{deluxetable*}
\end{center}

At 5 GHz, the map shows the {\em{ridge}} plus two-shell system already revealed by previous cm observations. However, the East shell is 
only partially contained in the field of view. The {\em{ridge}} and the West shell are 
also visible at 19 GHz. At this frequency, the circular shell structure appears almost complete, with the exception of an opening 
towards the East. Both the 5 and 19 GHz maps present filament-like structures, as well as regions of increased emission along the {\em{ridge}} and on the Southern part of the 
West shell. These knots are rather compact and their size is typically of the order of 10''. Similar knots also characterize the map at 34 GHz. 
Noteworthy, the 5 and 19/34 GHz maps reveal morphological differences in correspondance of the {\em{ridge}}. In particular, at 19 GHz and 34 GHz the maps present 
a few emission features which are either completely absent or only barely visible in the 5 GHz map, despite the fact that the size of these features 
is compatible with the range of angular scales which the 5 GHz map is sensitive to. For instance, the {\em{spur}} 
centered at (RA, Dec) = (10$^{h}$ 24$^{m}$ 16.2$^{s}$, -57$^{\circ}$ 45$^{\prime}$ 02''), 
roughly 40'' in size, is quite prominent and diffuse at 19 and 34 GHz, while it is rather faint and filamentary at 5 GHz (see Figure~2, 3 and 4). Likewise, the feature at 
(RA, Dec) = (10$^{h}$ 24$^{m}$ 18.0$^{s}$, -57$^{\circ}$ 46$^{\prime}$ 45.05''), $\sim$ 25'' long, is again bright and extended at 19/34 GHz, but
virtually invisible at 5 GHz (see Figure~2, 3 and 4). Such differences might be indicative of rising spectral indices or of an additional component of 
emission at high frequency. At the same time, they could be ascribed, at least in part, to the sensitivity to different angular 
scales of the individual data sets.

\subsection{Recombination line data} 

As mentioned in the previous section, the recombination line and continuum observations were 
performed simultaneously, thus covering the same 7.8$^{\prime}$ $\times$ 5.6$^{\prime}$ area of the sky. The 5 GHz recombination line 
data set was reduced again using CASA v3.3. The average noise level per channel was estimated to be of the order of $\sim$ 2.6 mJy/beam. 
Accordingly, line emission was found only at three positions (denoted as 'A', 'B', 'C') located  along the {\em{ridge}} 
(see Figure~2). We detected no emission in the inner part of the West shell. 

Using the task \texttt{viewer}, we extracted a cumulative spectrum within a box of the same size of the synthetic beam ($\sim$ 7'') and centered at each position. 
In our system configuration we had a continuum bandwidth of 13 MHz, which allowed us to 
estimate the continuum level, T$_{C}$, by fitting the data, minus the line, with a power-law model. A
gaussian fit was applied to the continuum-subtracted data (see Figure~5). Both the continuum and the line fits were carried out 
with the MATLAB fitting tool (\texttt{cftool}). The gaussian fit to the line gives the line peak temperature, $T_{L}$, central velocity, $V$, 
and width to half-intensity, $\Delta V$ (see Table~4).

We estimated the probability of a false detection. From Table~4, the statistical significance of the H109$\alpha$ line is $\sim$ 3$\sigma$ in region A 
and B, and $\sim$ 2$\sigma$ in region C, which corresponds to a probability of a spurious detection in a single velocity channel of, respectively, 1/370 and 1/50. Therefore, the probability that 
at least one out of N velocity channels has a S/N of 3 (or 2) is given by 1 - (1 - 1/370)$^{N}$ (or 1 - (1 - 1/50)$^{N}$). Since 
we apply the positivity condition{\footnote{The positivity condition states that an emission line - such as a RRL - always consists in a positive signal, while a noise 
spike can be both positive or negative.}}, the previous expression becomes 1 - (1 - 1/740)$^{N}$ (or 1 - (1 - 1/100)$^{N}$). We now consider the number of velocity  
channels that are effectively used for the detection of the line. First of all, the channels at the edge of bandwidth are discarded, as they are too noisy 
to provide any useful information. This operation brings the number N of channels from 512 to 400. In addition, we restrict the search of the line to the velocity range +/- 30 km/s, which  
is the range of velocities compatible with the position of RCW 49 in the l-v diagram (see Figure~2 of Furukawa et al. 2009). 
This further decreases N from 400 to roughly 40 channels. Following these considerations, 
we finally estimate that the probability that the lines in region A and B are spurious is 1 - (1 - 1/740)$^{40}$ $\sim$ 0.05, while for region C is 1 - (1 - 1/100)$^{40}$ $\sim$ 0.33. 
This calculation shows that our detections, especially in region C, are only marginal.

From the recombination lines, we computed the electron temperature, T$_{e}$, by applying the 5-GHz specific relation (Caswell $\&$ Haynes 1987)

\begin{equation}
T_e = 10150\left(\frac{T_L \Delta V}{T_C}\right)^{-0.87} \hspace*{0.5truecm} K
\end{equation}

The expression above is an approximation of the Shaver et al. (1983) formula (see their Eq.~(1)), 
with $\nu$ = 5 GHz and where we assume LTE conditions and that the fraction of ionized He is negligible ($n_e = n_{{H}^+}$). 

The emission measure, EM, is defined as 

\begin{equation}
EM = \int_0^\infty {n_e^2 dl}   \hspace*{0.5truecm} pc \hspace*{0.1truecm} cm^{-6}
\end{equation}

and it is derived from the 5-GHz continuum measurement. In fact, in the optically thin regime

\begin{equation}
T_{C} = T_{e} (1 - e^{-\tau_{C}}) \simeq T_{e} \tau_{C}   \hspace*{0.5truecm} K
\end{equation}

with $\tau_{C}$ the optical thickness for the continuum emission given by 

\begin{equation}
\tau_{C} = 8.24 \cdot 10^{-2} T_{e}^{-1.35}\left(\frac{\nu}{GHz}\right)^{-2.1} EM 
\end{equation}

From Equation~(2), if n$_{e}$ is constant, it also follows that 

\begin{equation}
n_e \simeq \sqrt{\frac{EM}{l}}   \hspace*{0.5truecm} cm^{-3}
\end{equation}

with $l$ the linear size of the emitting region. Table~4 provides the derived values of T$_{e}$, EM and n$_{e}$ by taking  
$l$ = 1 pc. Uncertainties on these quantities were obtained by propagating the errors on $T_{L}$, $T_{C}$ and $\Delta V$. We notice a significant 
spread in T$_{e}$ for the three positions for which line extraction has been performed i.e, from $\sim$ 4000 K in region 'C' to $\sim$ 16 000 K 
in region 'B', indicating large variations in the gas physical conditions along the {\em{ridge}}. The average T$_{e}$ is 8468 $\pm$ 6326 K, which is consistent with  
the value reported by Caswell $\&$ Haynes (1987) of 7300 K. Our cumulative spectra are characterized by a mean velocity of 9.8 $\pm$ 10.7 km/s and by a mean  
velocity width of 8.2 $\pm$ 4.7 km/s, which is is in agreement with the H137$\beta$ line observations by Benaglia et al. (2013). However, Caswell $\&$ Haynes (1987) and 
Churchwell et al. (1974) detect line emission for the H109$\alpha$ and the He109$\alpha$ line, respectively, centered at 0 and -4 $\pm$ 1 km/s and with a line width 
of 46 and 50 km/s. We can ascribe this apparent discrepancy to the presence of 
several velocity components along the {\em{ridge}}, possibly reminiscent of independent structures which happened to collide at some point in time, as recently 
proposed by Furukawa et al. (2009) based on their NANTEN2 CO(J = 2-1) observations.

The remaining results, concerning the EM and the electron density, will be discussed in Section~4.3.

\begin{figure}
\label{fig:line_ex}
 \begin{center}
 \includegraphics[width=0.85\linewidth]{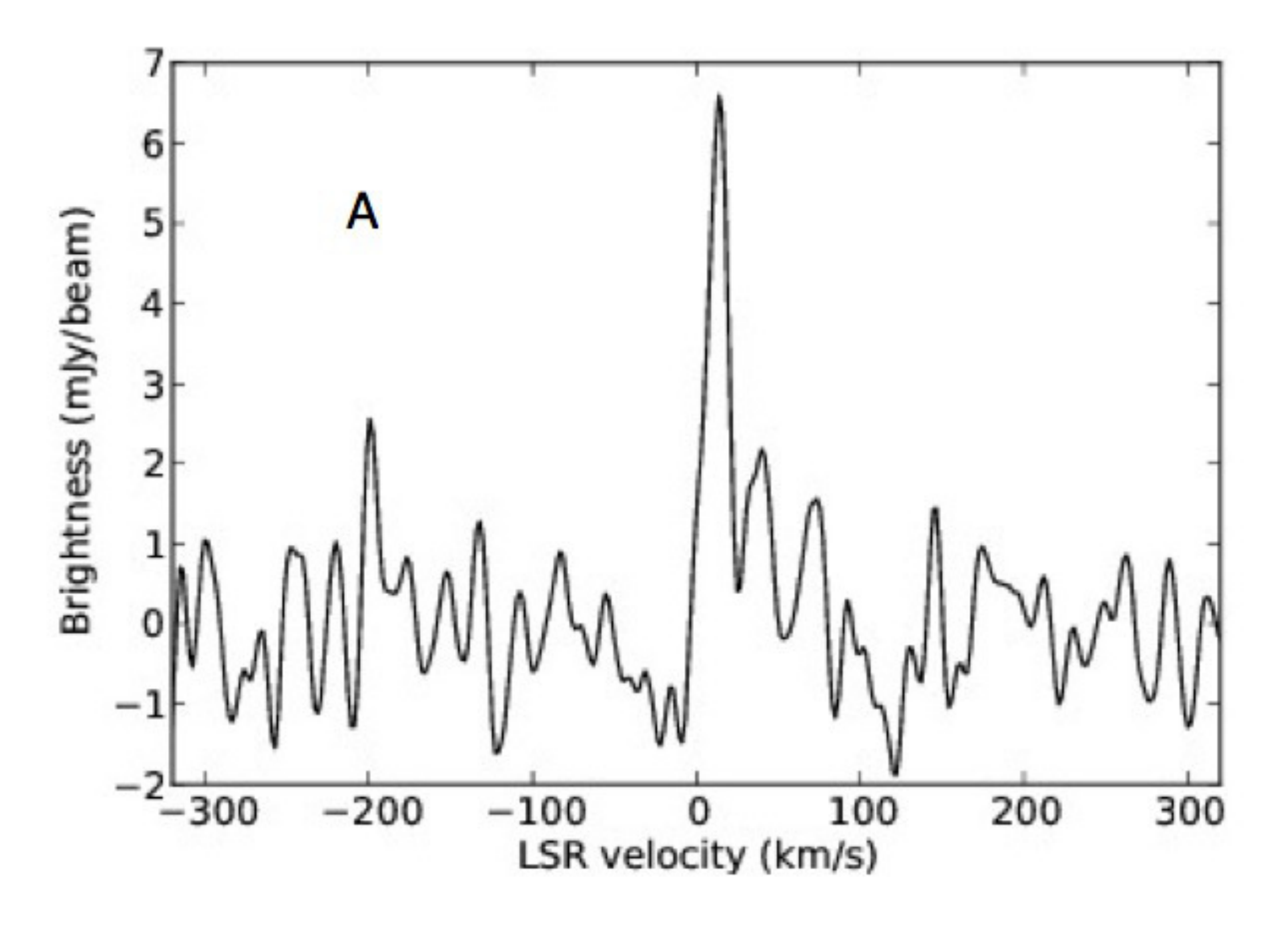}\\
 \vspace*{0.3truecm}
  \includegraphics[width=0.9\linewidth]{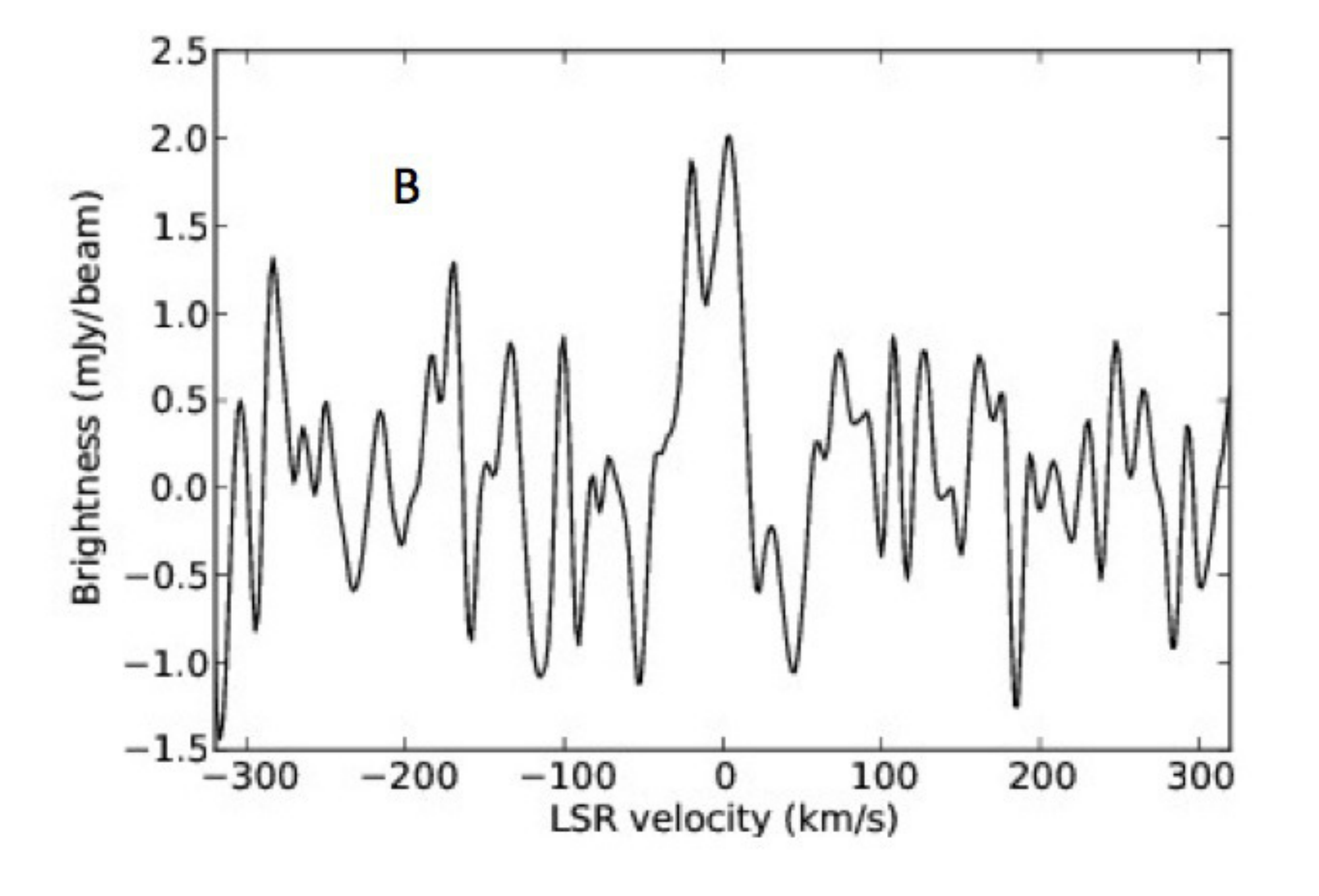}\\
 \vspace*{0.3truecm}
 \includegraphics[width=0.95\linewidth]{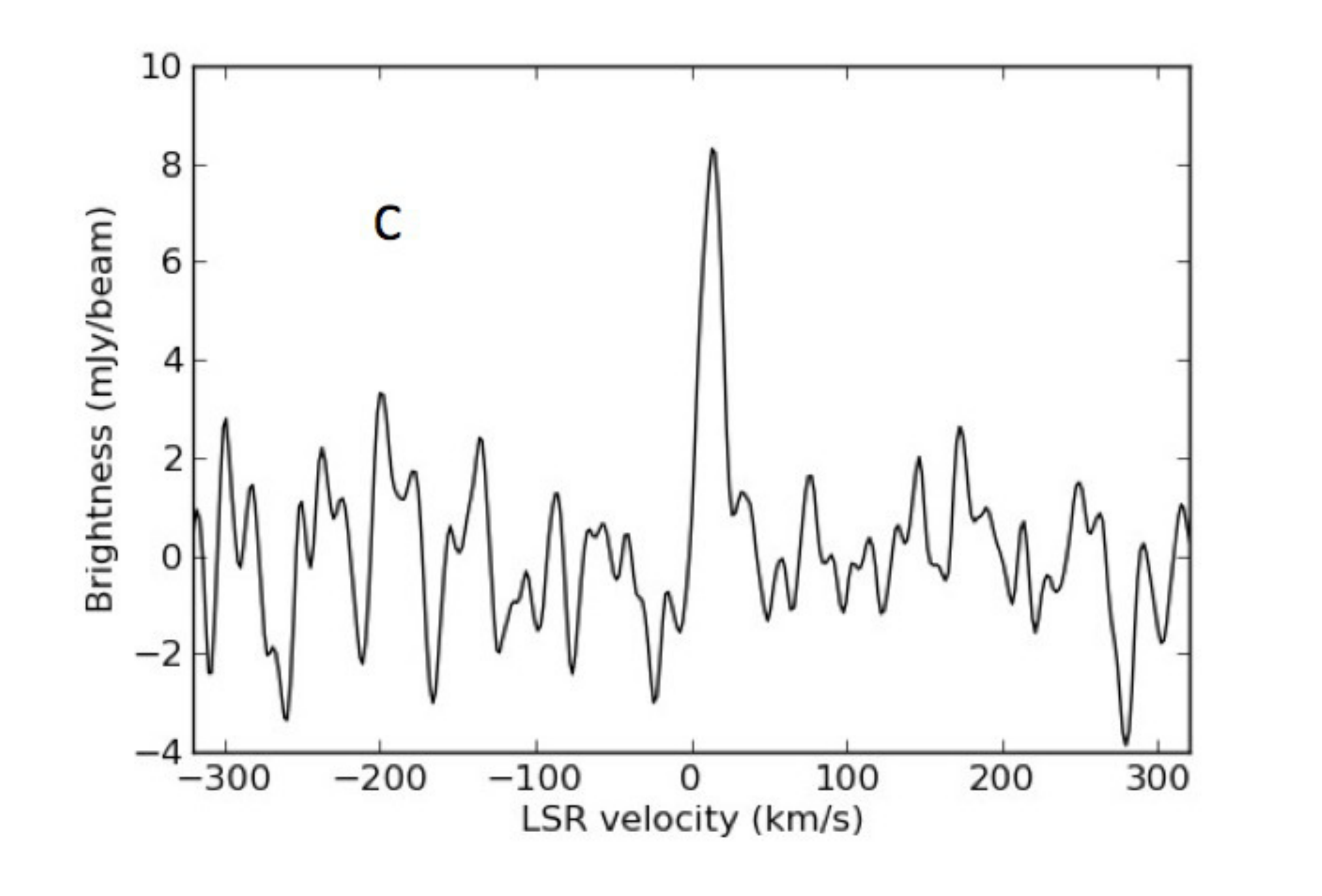}\\
 \caption{Cumulative continuum-subtracted spectra extracted from positions 'A' (top panel), 'B' (middle panel) and 'C' (bottom panel). The spectra have been 
smooothed using a 10-channel Hanning filter. The H109$\alpha$ line is centered on the LSR $V \sim$ 0 km/s.}
\end{center}
\end{figure}

\section{Analysis and discussion}

\subsection{Comparison with IR data}

As noted in the Introduction, the currently most accredited interpretation for the excess of emission at microwave frequencies is of electric dipole radiation
from spinning PAHs and/or VSGs, which entails that the emission at 20 - 60 GHz should be always found to spatially correlate with mid-IR
emission. Indeed, in their paper, Whiteoak \& Uchida (1997, hereafter WU97) remark that the 12/25 $\mu$m IRAS and the ATCA 1.4 GHz data appear to trace common structures, such as the two shells
and the {\em{ridge}}, and that moving towards longer wavelengths (i.e. 100 $\mu$m), this close correspondence significantly decreases. 

We already mentioned that in the core of RCW 49 the Spitzer MIPS 24 $\micron$ data and the WISE 22 $\micron$ data are fully saturated. However, this is 
not an issue for our analysis. In fact, as discussed by various authors (e.g. Paladini et al. 2012; Everett $\&$ Churchwell 2010), in HII regions the emission around 20 $\mu$m is likely 
associated to Big Grains rather than VSGs, contrary to what typically occurs in the Interstellar Medium (ISM). Therefore, to investigate the possible existence of an IR-microwave correlation in RCW 49, 
we compared the ATCA continuum data at 5, 19 and 34 GHz with the IRAC 8 $\mu$m observations from the GLIMPSE survey. 

The comparison was performed both in the $uv$-plane and in real space. For this purpose, we used the \textsc{casa} \texttt{simulator} tool which, 
at each observed ATCA frequency, allowed us to generate the corresponding visibilities for the IRAC 8 $\mu$m data set by taking into 
account the specific configuration of the interferometer. In particular, in order to obtain a perfect correspondence between observed 
and simulated $uv$-plane coverage, we used the combined \texttt{sm.predict} and \texttt{sm.openfromms} functions. As described in Section~2, the ATCA 5 GHz data set consists 
of a single pointing. In this case, to perform the simulation, a correction for the ATCA primary beam ($\sim$ 570'') was applied by multiplying  
the IRAC 8 $\mu$m data with a Guassian profile, centered at the pointing phase center. On the contrary, the ATCA 19 and 34 GHz maps were obtained with a mosaic 
observing strategy. Therefore, a simulation was performed for each pointing and the correction for the primary beam was applied using a 
Gaussian mask of 150'' and 80'', respectively. Figure~6 shows the simulated vs observed visibilities at each frequency. In these plots, in order to minimize the correlations in the 
$uv$-plane, we binned the $u,v$ cells in 10-min bins at 5 GHz and 1-min bins at 19 and 34 GHz, where the difference in bin width reflects the different integration time 
at the various frequencies. We computed the Pearson correlation coefficient{\footnote{The Pearson correlation coefficient provides an indication of the quality of a least squares fitting to the data. The coefficient 
can take values from +1 to -1, where +1 indicates a total positive correlation, 0 indicates no correlation, and -1 indicates a total negative correlation.}}, 
$\rho$, and obtained:  $\rho_{8,5}$ = 0.936 $\pm$ 0.002, $\rho_{8,19}$ = 0.848 $\pm$ 0.002, $\rho_{8,34}$ = 0.783 $\pm$ 0.009, indicating that 
the correlation strength slightly weakens as we go from 5 to 19/34 GHz. Note that the correlation errors have been computed from the binned data, in linear scale, 
taking into account the number of data points ($N^{-1/2}$) and their standard deviation.

\begin{figure}
 \begin{center}
    \includegraphics[width=0.9\linewidth,angle=0]{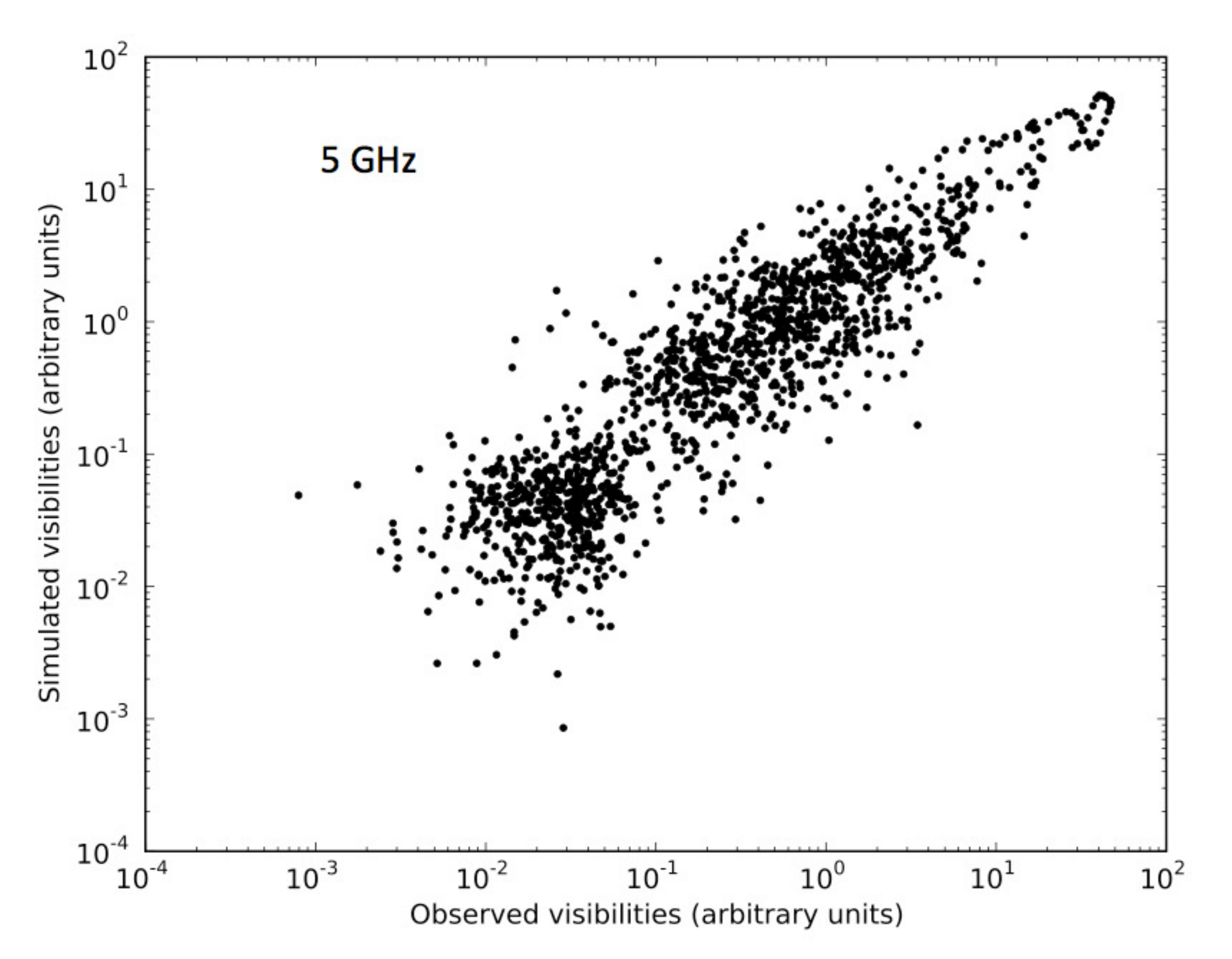}\\
  \vspace*{0.2truecm}
    \includegraphics[width=0.9\linewidth,angle=0]{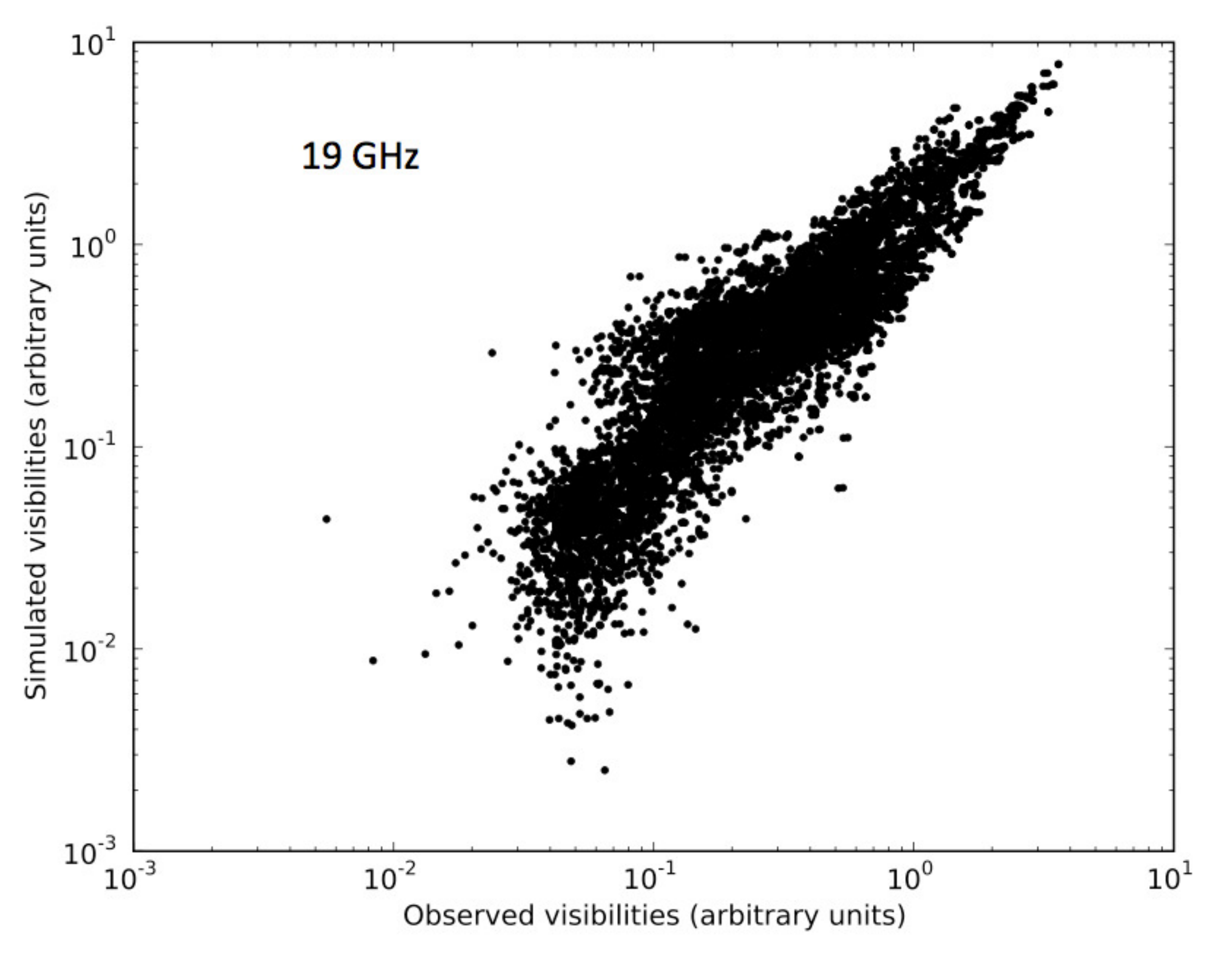}\\
 \vspace*{0.2truecm}
   \includegraphics[width=0.9\linewidth,angle=0]{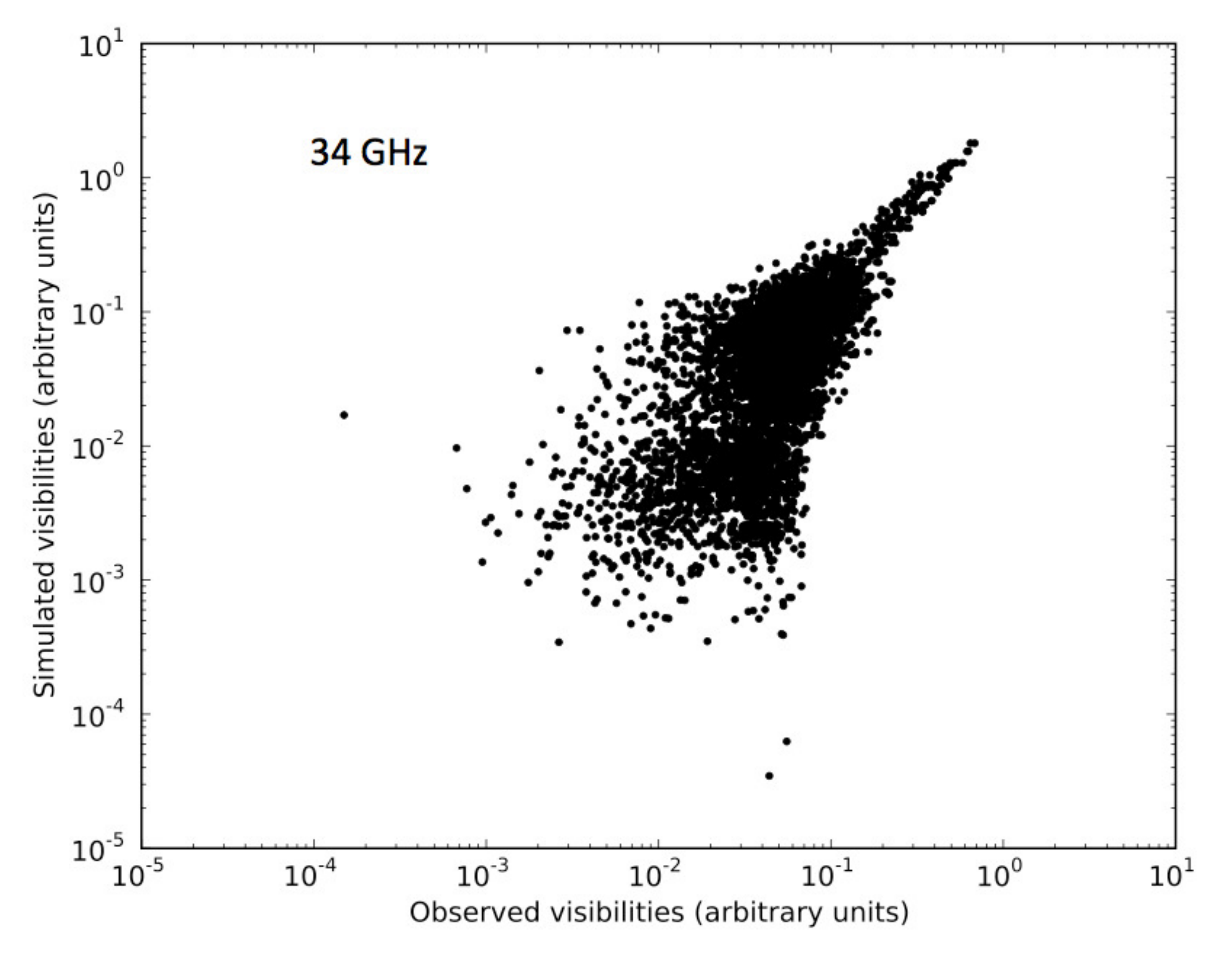}\\
 \vspace*{0.3truecm}
 \caption{The simulated IRAC 8 $\mu$m vs. the real ATCA visibilities. At 5 GHz (top panel), 
we notice a hint of a possibly residual correlation at high values of the visibilities.}
\end{center}
\end{figure}

The simulated IRAC 8 $\mu$m visibilities were then imaged to allow veryfing the robustness of the IR - radio correlation in real space. To minimize potential biases 
arising from the imaging process, we re-created the observed ATCA maps by using the same deconvolution parameters as for the simulated data set. In particular, 
all the observed maps were re-generated using the same H\"ogbom algorithm and the multiscale cleaning method (Cornwell 2008), with three 
windows of 0, 5 and 15 pixels, which we used for the IRAC simulated maps. The pixel size of all maps was set to 1.2'', and to each map 
the primary beam correction was applied. Figure~7 illustrates, for the case of the  {\em{ridge}}, a morphological 
comparison between the simulated IRAC 8 $\mu$m and the observed ATCA 5, 19 and 34 GHz data. 
At 5 GHz (top panel), both the 8 $\mu$m and radio continuum emission 
peak in the South-East part of the {\em{ridge}}, although clear descrepancies can be noticed between the two emission components. Indeed, radio 
emission fills a bubble-like structure located to the North-East of the {\em{ridge}}, for which no 8 $\mu$m emission is detected. Such differences become 
even more evident at 19 and especially at 34 GHz (middle and bottom panels): while the 8 $\mu$m and microwave emission peaks still spatially coincide towards 
the East side of the {\em{ridge}}, the West side is characterized, in the cm maps, by a prominent  
shell-like structure and by a spur with no IR counterpart. These shell-like structure and the spur are the same as discussed in Section~3.1. 

\begin{figure}
 \begin{center}
  \includegraphics[width=0.9\linewidth]{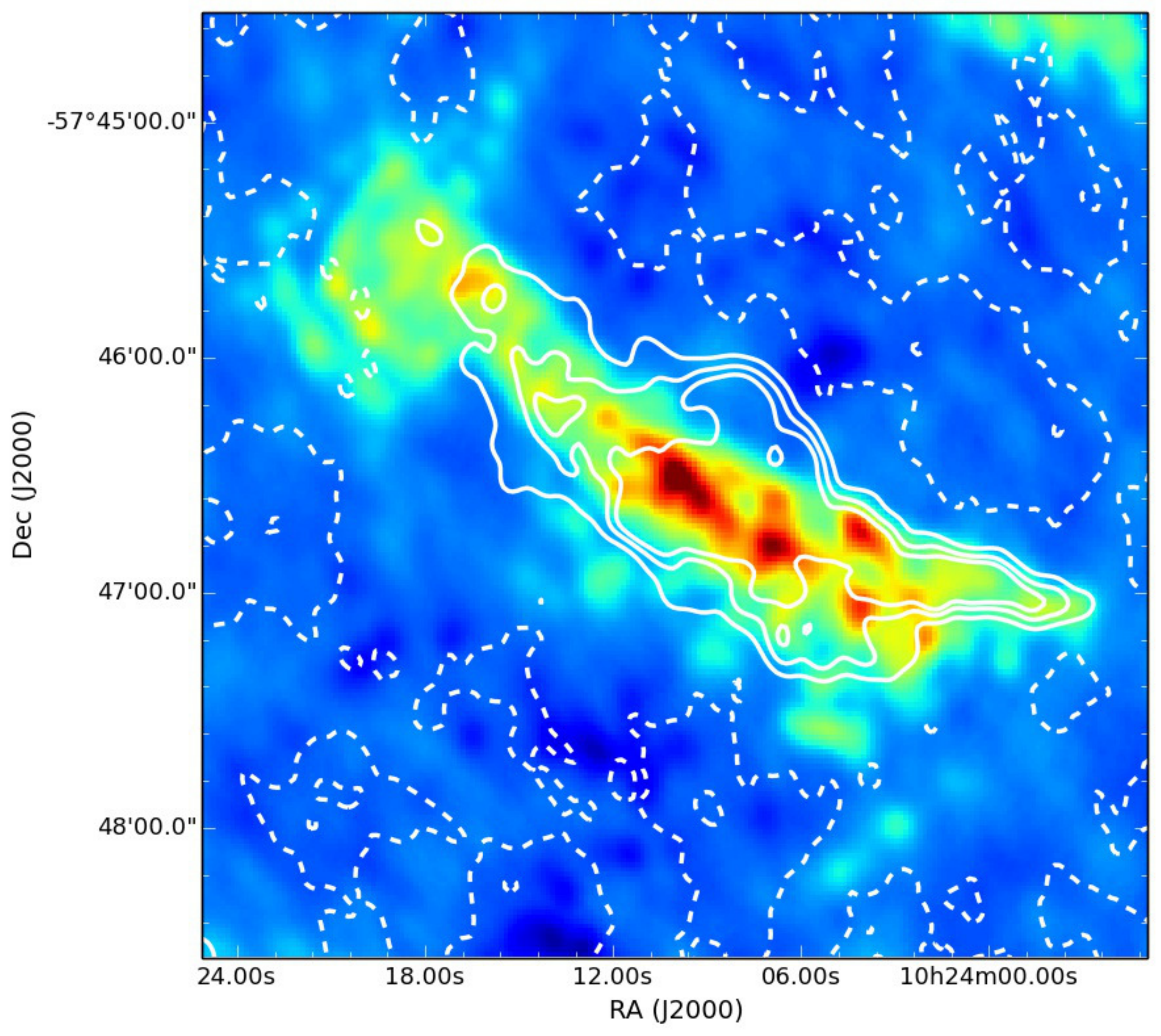}\\
  \includegraphics[width=0.9\linewidth]{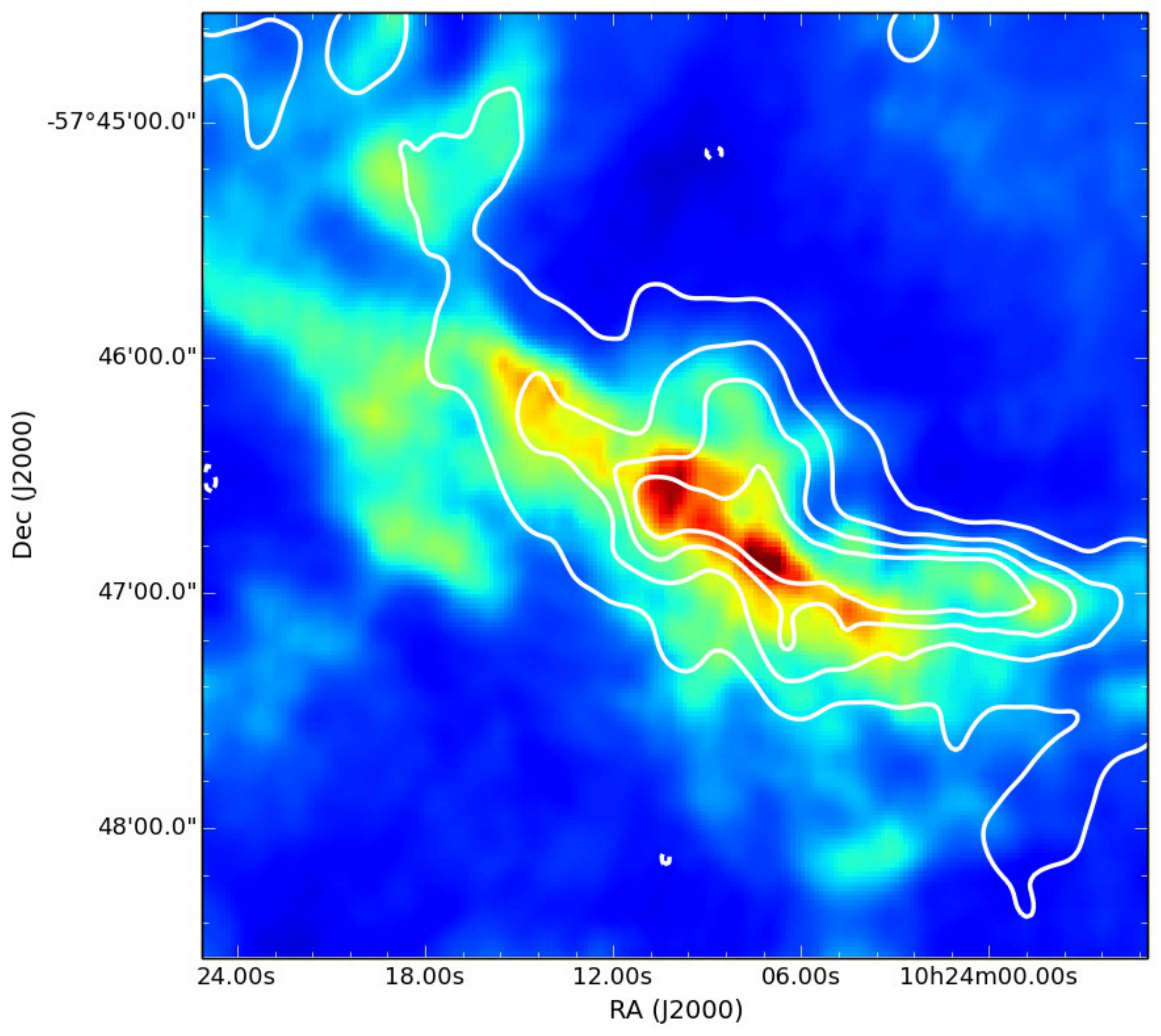}\\
  \includegraphics[width=0.9\linewidth]{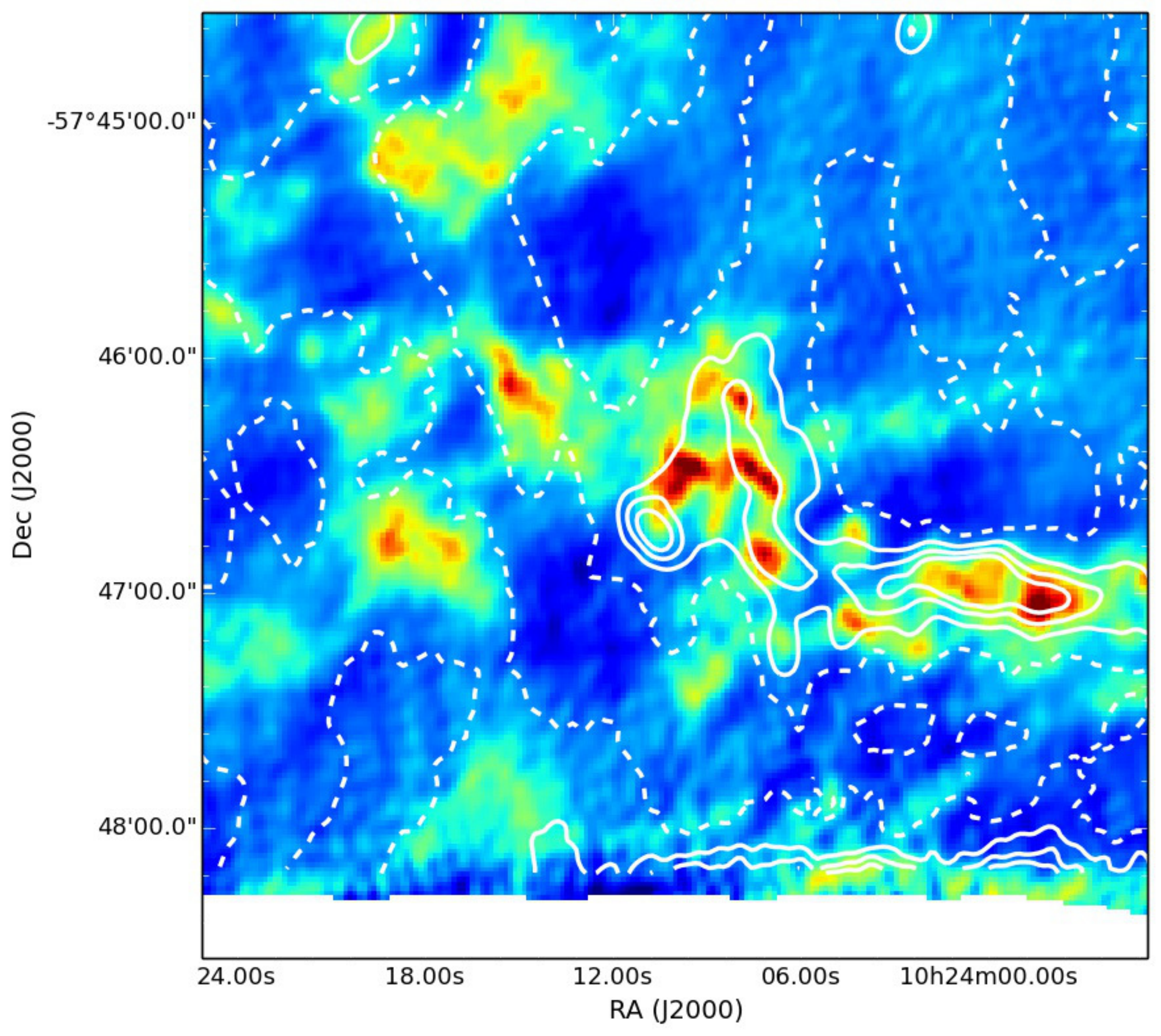}\\
 \caption{Simulated IRAC 8 $\mu$m observations (white contours) overlaid on the observed ATCA data. From top to
bottom: 5 GHz (750 B configuration), 19 GHz and 34 GHz (EW 352 configuration).}
\end{center}
\end{figure}

\subsection{Spectral index map} 

Ideally, we should estimate the amount of microwave excess in the ATCA data by extrapolating the low-frequency
observations (5 GHz) to the high-frequencies (19 and 34 GHz) and then by comparing the result with the CBI 31 GHz measurement. 
Since different interferometer configurations correspond to different amounts of flux loss, a correct 
approach to this problem would entail filtering the data in the $uv$-plane to assure  
that, at each frequency, we are sensitive to emission on the same angular scales. However, given the poor overlap in 
the visibility plane of the ATCA data set, if we followed
this strategy we would remain with virtually no data to analyze.  

To circumvent this problem, we considered only low frequency data and, in particular, a combination of our new 5 GHz data 
and of archival ATCA 1.4 GHz data from WU97. These two data sets are characterized by sufficient overlap in the $uv$-plane 
to allow the generation of a reliable spectral index map. To this end, the $uv$- range of the
1.4 GHz data was truncated to match the shortest $uv$-coverage of the 5 GHz
data (e.g. 1.1 k$\lambda$). In addition, since the 5 GHz data still have a denser $uv$-coverage at short spacings, we truncated these to the same 
maximum $uv$-distance of the 1.4 GHz dataset (e.g. 20.55 k$\lambda$), in order to guarantee
a similar shape of the weighting function during the imaging process. It is important to emphasize that this
operation does not lead to matched beams at 1.4 GHz and 5 
GHz, but simply mitigates the influence of extended structures at 1.4 GHz, and
avoids underweighting of the short spacings at 5 GHz. After cleaning, the 1.4 GHz and 5 GHz maps were restored with the same
synthetic beam (15''). The maps were then converted in
Jy/pixel units, and pixels below 3 times
the standard deviation computed for each map were discarded. Finally, for each pixel the spectral
index (S$_{\nu}$ $\sim$ $\nu^{\alpha}$) was computed using the relation:

\begin{equation}
 \alpha=\frac{log(S_{5 \hspace*{0.1truecm} GHz}/S_{1.4 \hspace*{0.1truecm} GHz})}{log(5/1.4)}
\end{equation}

Figure~8 shows the spectral index map, and its associated uncertainty map, obtained following
the aforementioned steps. The noisy behaviour, especially at the
edges of the central stucture (the {\em{ridge}}) and
of the two shells, likely derives from small differences of the synthetic beams at 1.4 and 5 GHz. Note that emission contribution on specific spatial scales 
might differ pixel to pixel, introducing a bias on the reconsctructed spectral
index map which is difficult to quantify.

Although we cannot directly estimate the uncertainty associated with the interpolation of the sampling function, it is commonly accepted that 
this is a function of the $uv$-weigths, hence of the noise. 
Therefore we assume that the error map obtained by considering the average rms is a good representation of the uncertainty of the spectral index map. 
In particular, we determined the error map in each pixel by propagating Eq.~(4), i.e., as:

\begin{equation} 
 \Delta\alpha=\frac{\sqrt{\left(\frac{\sigma(5\,\rm GHz)}{S(5\,\rm GHz)}\right)^2+\left(\frac{\sigma(1.4\,\rm GHz)}{S(1.4\,\rm GHz)}\right)^2}}{\log(\frac{5}{1.4})} 
\end{equation} 

where $\sigma$ was evaluated at different positions of the maps and then averaged. We emphasize that the above relation does not take into account modelling errors 
and/or systematics, which are difficult to quantify. 
As shown in Figure~8, right panel, the uncertainty in the central part of the {\em{ridge}} and in a large part of the southern shell varies between +0.05 and +0.1, 
while at the edges of the emission regions it increases up to about +0.3. 

\begin{figure*}
\label{fig:sp_map}
 \begin{center}
  \includegraphics[width=0.47\linewidth]{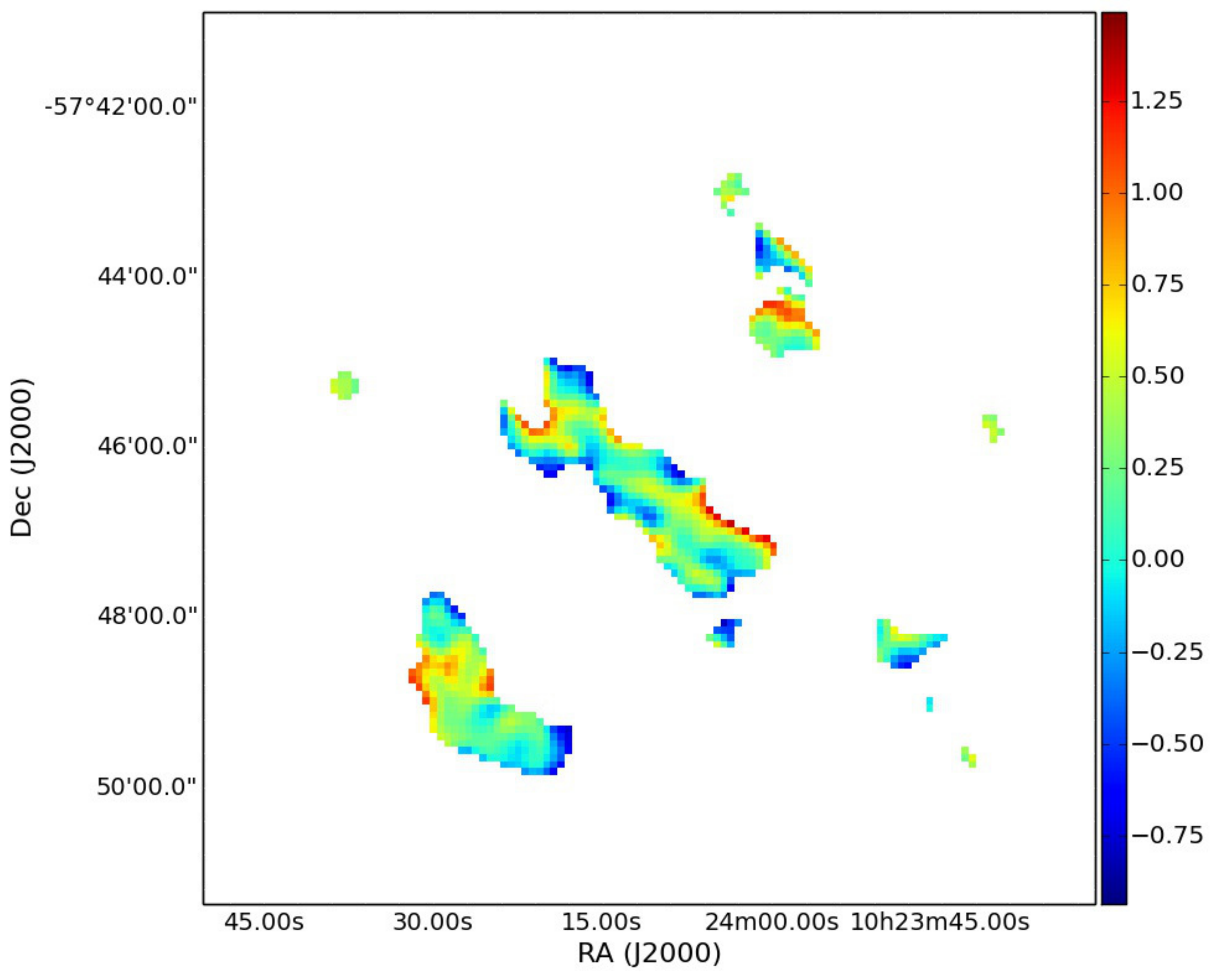}
  \includegraphics[width=0.47\linewidth]{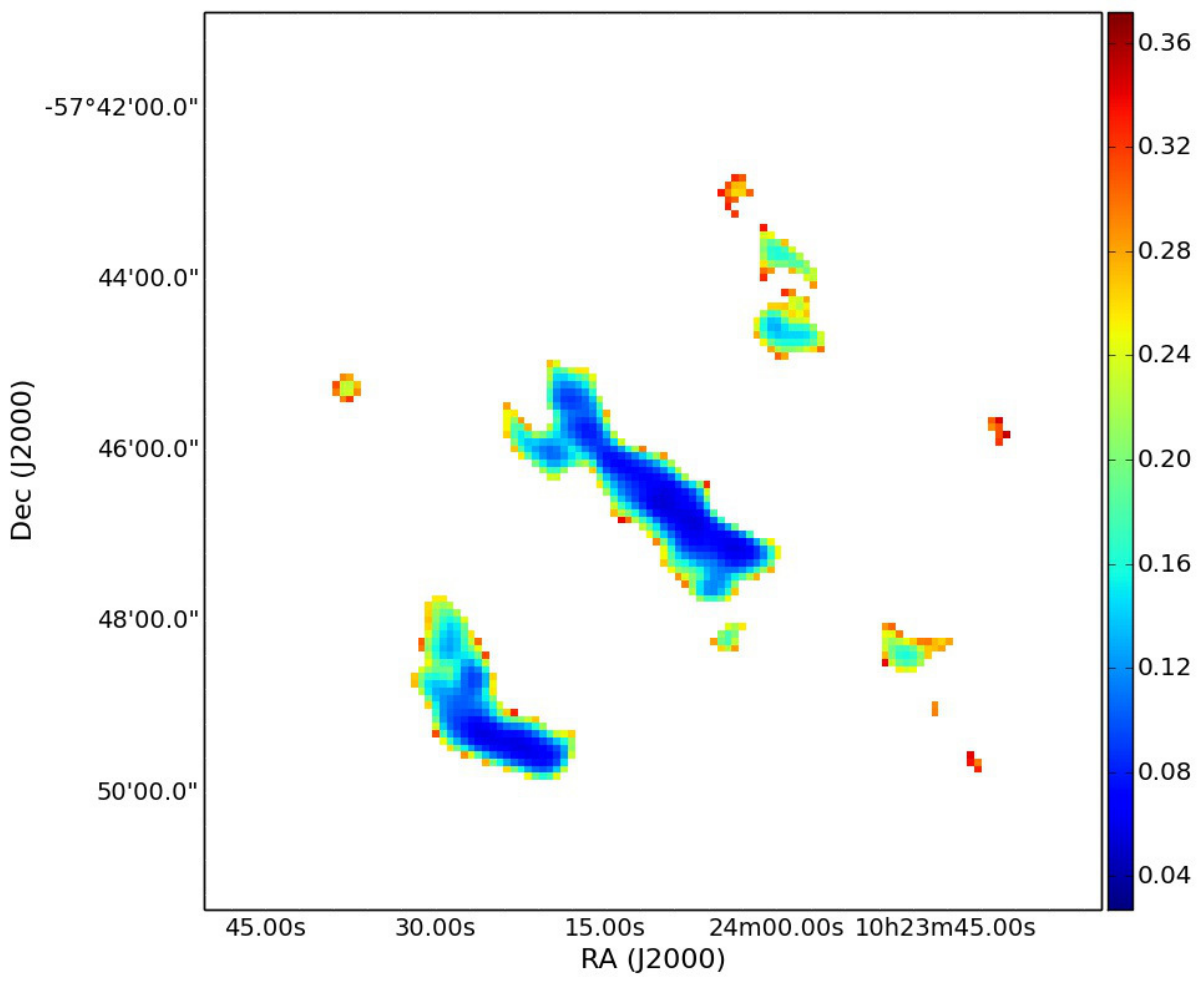}
 \caption{Left: spectral index map obtained by combining the 5 GHz new observations of the core of RCW 49 with ATCA archival data at 1.4 GHz. 
The S$_{\nu}$ $\sim$ $\nu^{\alpha}$ convention is used. Right: spectral index error map.}
\end{center}
\end{figure*}

If we exclude the noisy pixels (i.e., those with an associated uncertainty greater than 0.1), the mean of the spectral 
index distribution is +0.20 $\pm$ +0.22. This is consistent (within 1.5$\sigma$) with optically thin free-free emission ($\alpha$ = -0.1), likely due to the ionizing flux from the OB stars in the 
Wd2 cluster. However, the dispersion around the mean is quite significant. In particular, both the {\em{ridge}} 
and the West shell are characterized by positive spectral indices. Benaglia et al. (2013) have recently published two spectral index maps of RCW 49: the first one, derived    
from the combination of archival ATCA 1.4 and 2.4 GHz data{\footnote{The 1.4 GHz data set is the same one from WU97 used in this work.}}, 
covers both the {\em{ridge}} and the shells, while the second, generated from newly obtained 5.5 and 9.0 GHz data, covers only the {\em{ridge}}. 
The 1.4 -- 2.4 GHz spectral index map is sensitive to angular scales up to several arcmins. The 5.5 -- 9.0 GHz map is sensitive only up to scales of the order of 100'', that is a  
smaller scale than the maximum one which our 1.4 -- 5.0 GHz map is sensitive to ($\sim$ 3.4'). The spectral index map obtained 
from the lower frequency data (their Figure~5) appears to be fully compatible with optically thin free-free emission. This map also shows 
a possible contribution from synchrotron emission, especially along the Northern side of the {\em{ridge}} and in the West shell. Conversely, the spectral index map 
associated with the higher frequency data (their Figure~6) shows positive spectral indices. Indeed a large fraction of the {\em{ridge}} appears to be 
characterized by spectral indices greater than +0.3, in agreement with our analysis. Interestingly, values of the order of +0.6 are 
typically associated with stellar winds (e.g., Panagia $\&$ Felli 1975) and, in general, a global distribution with rising spectral indices 
might corroborate the hypothesis, advanced by WU97, that the {\em{ridge}} is the product of the direct collision   
between the high-velocity winds produced by WR20a and WR20b in the two shells or, alternatively, of a collision between the two shells
before merging. Remarkably, this collision scenario could also explain the increase of dust temperature along the {\em{ridge}} noted by 
the same authors.

\subsection{No evidence for self-absorbed free-free emission}

As we discussed in Section~2, one of the main objectives of the ATCA observations was to assess the presence, in the core of RCW 49, of
compact high-density components which could account for an inverted emission spectrum at microwave frequencies. In the Altenhoff et al. (1960)
approximation, the free-free optical thickness, $\tau_{ff}$, is a function of T$_{e}$, $\nu$ and EM according to:

\begin{equation}
\tau_{ff} \simeq 0.08235 \hspace*{0.1truecm} T_{e}^{-1.35} \hspace*{0.1truecm} \nu_{GHz}^{-2.1} \hspace*{0.1truecm} EM
\end{equation}

where T$_{e}$ is in K, $\nu$ in GHz and EM in pc cm$^{-6}$. For T$_{e}$ of the order of 8000 K,
we expect $\tau_{ff} >$ 1 at 31 GHz, which is the frequency of the CBI tentative detection of AME in RCW 49, 
when EM is greater than $\sim$ 10$^{9}$ pc cm$^{-6}$. These extremely high emission measures are typical of hyper-compact
HII regions (HCHII), which are also characterized by high electron densities (n$_{e} >$ 10$^{6}$ cm$^{-3}$) and
linear diameters of less than $\sim$ 0.05 pc. The very small size of these objects make them good candidates for being
responsible for the emission observed by the CBI while remaining virtually undetected in its low-resolution
beam. Moreover, the existence of HCHIIs in the core of RCW 49 would be consistent with a triggering scenario, which is known to
be common in Galactic HII regions (Elmegreen 1998; Deharveng et al. 2005). It is remarkable, from this point of view, that the GLIMPSE observations
of RCW 49 have indeed allowed the identification of
some 300 likely young stellar objects (YSOs), representing a new generation of star formation, possibly triggered by the stellar winds and
shocks associated with the Wd2 cluster (Whitney et al. 2004). This YSO population is found in six different regions spread across RCW 49,
none of which coincides with the 31-GHz CBI emission peak. The closest region (denoted region '3' by Whitney et al. 2004) is situated to the South-West of the CBI peak, 
and at a distance of $\sim$ 2'.

From the analysis of the H109$\alpha$ data set described in Section~3.2, we estimated the emission measure and
electron density at three different positions along the {\em{ridge}}. We recall that no line emission was found, above the sensitivity level, 
at any other locations, although we should emphasize that the thermal continuum emission also encompasses only $\sim$ 1/20$^{th}$ of the 7.8$^{\prime}$ $\times$ 5.6$^{\prime}$ 
covered area (roughly 147 beams out of 2950).

Table~4 shows that the average EM is $\sim$ 160 pc cm$^{-6}$, with a
corresponding n$_{e}$ of only $\sim$ 13 cm$^{-3}$. These values are indicative of ionized gas denser than in typical ISM conditions, for which
0.03 $<$ n$_{e} <$ 0.08 cm$^{-3}$ (Haffner al. 2009), supporting the compressed-by shocks hypothesis formulated by WU97, yet this same gas appears to be much more diffuse than 
in HCHIIs. From the non-detection of H109$\alpha$, we can set lower limits for 
the electron temperature and emission measure through most of the 5 GHz map. For this purpose, we assume as representative of the
non-detection region, T$_{c}$ $\simeq$ 60 mJy/beam and $\Delta$V = 8.2 km/s. The latter is the average value for the H109$\alpha$ line in regions A, B and C. 
Setting the detection threshold at 3$\sigma$, where $\sigma$ = 2.6 mJy/beam, and using Equation~(1), (2), (3) and (4), we obtain T$_{e}$ $>$ 9740 K, EM $>$ 376 pc cm$^{-6}$, 
and n$_{e}$ > 19.4 cm$^{-3}$. Therefore our RRL measurements cannot completely rule out the presence of HCHIIs in the core of RCW 49. However, considering 
that the spectral index map derived in Section~4.2 shows no evidence of self-absorbed free-free emission ($\alpha$ $\sim$ 2), which would certainly characterize 
such dense condensations of ionized material, we confidently state that it is unlikely that the CBI detection can be attributed to a hidden population of HCIIs.

\section{Discussion}

The main limitation to keep in mind in interpreting our results is the fact that the ATCA is an interferometer working at arcsec resolution and, as such, it is 
subject to significant flux loss on angular scales larger than a few arcmins. In our case, this means that, although 
the area imaged with the ATCA is 7.8$^{\prime}$ $\times$  5.6$^{\prime}$, at 5, 19 and 34 GHz, 
the emission on angular scales larger than 3.4', 1.7' and 1', respectively, is filtered out. The total 31-GHz flux detected by CBI towards RCW 49 in a 6.8' beam 
is 146.5 $\pm$ 5.2 Jy. 
If we integrate the ATCA 5 GHz flux in a similar aperture, we obtain 10.3 Jy which, when extrapolated to 31 GHz using the mean 
spectral index value (+0.20 $\pm$ +0.22) derived in Section~4.2, gives 14.82 Jy, that is $\sim$ 10$\%$ of the CBI observed flux. A similar calculation at 
19 and 34 GHz provides, respectively, 31 GHz projected fluxes of 10.18 Jy and 3.11 Jy corresponding to 7$\%$ and 2$\%$ of the CBI flux. These estimates clearly 
show that, because at all ATCA frequencies we are losing an important fraction of flux, we have to be careful in generalizing the results of 
our analysis, especially those regarding the spectral indices. However, it is still possible that, due to the uncertainties inherent to the filtering process in 
the $uv$-plane, more than the 30$\%$ of the spectral indices are actually positive and not consistent with the canonical value for optically thin free-free emission. In the hypothesis 
where this rising can be ascribed to a collission between shells or stellar winds, it is hard to imagine that such an energetic phenomenon would affect 
the emission only on small angular scales. Alternatively, we can speculate that the rising that we observe between 1.4 and 5 GHz is already indicative 
of AME at these low frequencies. If this was the case, the implication would be that PAHs with a larger size range than typically adopted by spinning 
dust models (3.5 to 100 \AA, e.g. Ali-Ha{\"imoud} et al. 2009) are likely contributing to the emission. 

One important thing to notice is that in RCW 49 the location of the CBI peak is offset ($\sim$ 50'') with respect to the peak of dust emission (see Figure~1). 
This is also the case of RCW 175, for which Tibbs et al. (2012) have shown, by means of 
a detailed modelling analysis of the dust content of the region, that the AME peak does not coincide with the peak of the PAH and VSG spatial 
distributions, and that such a shift cannot be attributed to instrumental effects e.g., the CBI pointing error (up to 30'') or its low angular resolution. 
This finding alone suggests that either the 
detected AME is not due to spinning dust or, if the origin of the emission is indeed spinning dust, other factors (such as gas ions, etc.) might 
play a pivotal role in triggering this type of emission, as discussed in Tibbs et al. (2012).

\section{Conclusions}

We presented the analysis of ATCA multi-frequency observations (5, 19 and 34 GHz) of a 7.8$^{\prime}$ $\times$  5.6$^{\prime}$ region centered 
on the core of RCW 49, the brightest HII region in the Southern hemisphere. At 5 GHz we observed both the continuum and the H109$\alpha$ hydrogen recombination line. 
Using this new ATCA data set, we investigated the potential correlation between the radio continuum emission at each observed frequency and the 8 $\mu$m emission, which 
is dominated by the 7.7 $\mu$m feature commonly attributed to PAHs. Our analysis shows that, overall, the correlation persists from short (5 GHz) to long wavelengths (34 GHz), and 
from large ($\sim$ 3.4') to small (0.4'') angular scales. However, we found an indication of a weakening of the correlation going to higher frequencies and to small angular scales. 
By combining, respectively, new and archival 5 GHz and 1.4 GHz ATCA data, we generated a spectral index map (and its corresponding error map) for the region, encompassing both the {\em{ridge}} 
and the shells. The spectral index distribution is globally consistent with optically thin free-free emission, although it is characterized by a large dispersion. In particular, $\sim$ 30$\%$, 
20$\%$ and 10$\%$ of the spectral indeces are greater than +0.3, +0.4 and +0.5, respectively, with an average error of less than +0.1. These positive indeces are found 
mostly along the South side of the {\em{ridge}} and of the West shell. This rising of the spectral index distribution might suggest that other mechanisms, 
other than spinning dust, are responsible for the microwave excess observed by CBI, such as shocks due to a collision 
event either between the West and East shells or between the stellar winds generated by WR20b and WR20a, as speculated by WU97. Finally, the combined 
analysis of the spectral index map with the H109$\alpha$ recombination line, 
marginally detected at three positions along the {\em{ridge}}, appears to suggest that the core of RCW 49 is likely not harboring 
very compact sources such as HCHII regions, thus not supporting an inverted free-free emission scenario as explanation for the CBI 31 GHz excess. 

We believe that this study has mostly shown the importance of obtaining high angular resolution observations even for sources which 
have been studied for more than a century, like HII regions. These observations are in fact key in revealing the complexity of physical mechanisms at work 
in these objects, in particular in very extended and evolved ones such as RCW 49. Without this detailed information it is 
difficult to interpret unambiguously the emission seen at relatively low angular resolution at microwave frequencies.

\acknowledgments
The authors would like to thank the anonymous referee for his/her constructive comments and Dr. Phil Appleton 
for useful discussions. CD acknowledges support from an STFC Consolidated Grant (ST/L000768/1) and an ERC Starting Grant (no.$\sim$307209).

%%% BIBLIOGRAPHY

%\bibliography{rcw49_atca_v2}

\end{document}